\documentclass[nofootinbib]{article}
\usepackage[utf8]{inputenc}
\usepackage[T1]{fontenc}
\usepackage{geometry}
\usepackage{amsmath,amssymb,amsfonts,bm}
\usepackage{graphicx}
\usepackage{braket}
\usepackage{authblk}
\usepackage{lettrine}
\usepackage{color}
\usepackage{times}
\usepackage{hhline}
\usepackage{tabularx}
\usepackage{calc}
\usepackage{soul}
\usepackage{mathbbol}
\usepackage{makecell}
\usepackage{subcaption}
\usepackage{multirow}
\usepackage{wrapfig}
\usepackage{xcolor}
\usepackage[table]{xcolor}
\usepackage{booktabs}
\usepackage{parskip}
\usepackage{siunitx}
\usepackage{pbox,cellspace}
\usepackage{colortbl}
\usepackage[numbers,sort&compress]{natbib}
\usepackage[colorlinks=true,linkcolor=blue,citecolor=blue,urlcolor=blue]{hyperref}

\renewcommand{\thefootnote}{\fnsymbol{footnote}}

\title{Gauge-covariant theory of the orbital Hall effect: 
\\
role of itinerant orbital currents and a non-Kubo contribution}

\author[1,2,3,*,$\dagger$]{Mirco Sastges}
\author[4,5,*]{Insu Baek}
\author[4,5,*]{Hojun Lee}
\author[4,5]{Hyun-Woo Lee}
\author[1,2,$\ddagger$]{Yuriy Mokrousov}
\author[5,6,$\S$]{Dongwook Go}
\affil[1]{Peter Gr\"unberg Institut, Forschungszentrum J\"ulich, 52425
J\"ulich, Germany}
\affil[2]{Institute of Physics, Johannes Gutenberg-University Mainz, 55099
Mainz, Germany}
\affil[3]{Department of Physics, RWTH Aachen University, 52056 Aachen, Germany}
\affil[4]{Department of Physics, Pohang University of Science and Technology,
Pohang, Kyungbuk 37673, South Korea}
\affil[5]{Center for Quantum Dynamics of Angular Momentum, Pohang 37673, South
Korea}
\affil[6]{Department of Physics, Korea University, Seoul 02841, South Korea}

\date{}

\begin{document}
\maketitle

\begin{abstract}
The orbital Hall effect is central to the emerging field of orbitronics, yet its first-principles description relies on an atom-centered approximation (ACA) of the orbital angular momentum (OAM) that discards itinerant contributions. Attempts to go beyond it have used off-diagonal elements of the ground-state orbital magnetization formula, which are not gauge covariant and treat the OAM as a fixed operator. Here we construct a gauge-covariant OAM operator in a Wannier basis that is consistent with the modern theory of orbital magnetization for all band pairs. Because this operator depends on the geometry of the Bloch states, an applied electric field modifies the operator itself, giving rise to a non-Kubo contribution to the orbital Hall conductivity (OHC). First-principles calculations for transition metals and monolayer MoS$_2$, which are frequently used in spin-orbitronics experiments, show that OHC contributions originating from the itinerant-circulation of the Wannier functions are often comparable to or larger than those originating from their local-circulation, which is especially evident in Pt and Fe. Separating the OHC instead into local and non-local terms, we find that the former closely resemble the ACA results, while the latter often dominate and reverse the sign predicted by the ACA. Our results call for a reassessment of how orbital currents are identified in experiments.
\end{abstract}

\footnotetext[1]{These authors contributed equally to this work.}
\footnotetext[2]{\hyperlink{mailto:m.sastges@fz-juelich.de}{m.sastges@fz-juelich.de}}
\footnotetext[3]{\hyperlink{mailto:y.mokrousov@fz-juelich.de}{y.mokrousov@fz-juelich.de}}
\footnotetext[4]{\hyperlink{mailto:dongwookgo@korea.ac.kr}{dongwookgo@korea.ac.kr}}

\renewcommand{\thefootnote}{\arabic{footnote}}

\section*{Introduction}

\quad There are two distinct sources of magnetism in matter: spin and orbital degrees of freedom~\cite{soderlind1992spin,eriksson1992spin}. While spin physics and the description of spin phenomena have been at the center of intensive research for decades, interest in orbital phenomena was sparked more recently~\cite{lopez2012wannier, go2021orbitronics,salemi2022first,jo2018gigantic}. This shift can be attributed to the realization that the argument of crystal-field quenching of orbital angular momentum (OAM) does not apply out of equilibrium~\cite{lopez2012wannier, go2021orbitronics,ando2025orbitronics,wang2025orbitronics}. As has been shown unambiguously in recent years, many solids exhibit field-induced orbital magnetism and transport that surpasses their spin counterpart by far~\cite{lopez2012wannier,ando2025orbitronics,wang2025orbitronics,go2024first}. In particular, this concerns the phenomenon of the orbital Hall effect (OHE), where electrically generated orbital currents in light materials are predicted to be orders of magnitude larger than the respective spin Hall currents~\cite{go2024first,salemi2022first,rang2025orbital,go2018intrinsic}. Consequently, the increasing interest in OAM and orbitronics underscores the importance of accurately assessing the OAM operator based on a rigorous, so far missing, framework with high predictive power.

A key to the predictive theory of orbital effects in complex materials is the accurate evaluation of the OAM operator in the basis of suitable quantum states~\cite{cysne2022orbital,pezo2022orbital,busch2023orbital,cysne2025orbital}. In solid-state theory, quantifying the OAM operator in terms of Bloch states presents a formidable challenge due to the unboundedness of the position operator~\cite{thonhauser2005orbital,go2024first}. The most widespread way to overcome this difficulty is the so-called atom-centered approximation (ACA), which is frequently regarded as an efficient means of assessing local, atomic-like orbital properties~\cite{hanke2016role,aryasetiawan2019modern}. While this approach has proved effective in many cases, for example when considering the interaction of orbital currents with localized spins in magnetic systems~\cite{ceresoli2010first,pezo2022orbital,hanke2016role}, it lacks the ability to clearly identify itinerant and non-local contributions to orbital dynamics. Alternative approaches that sought to quantify non-local OAM often failed to account for surface contributions or to preserve the gauge-covariance of their off-diagonal elements, resulting in an OAM operator that deviates from the rigorous theory of orbital magnetization~\cite{pezo2022orbital,bhowal2021orbital}.

In contrast to Bloch states, Wannier functions, which are localized in real space, have been shown to provide a viable solution to this problem, leading to the establishment of a practical Wannier-based philosophy for addressing various problems in solid-state physics, such as electric polarization~\cite{resta2010electrical,stengel2006accurate,noel2001polarization}, piezoelectricity~\cite{vanderbilt2000berry,noel2001polarization}, and the anomalous Hall effect~\cite{wang2006ab,xu2024maximally}. Notably, the Wannier picture of solids has been used to arrive at the modern theory of orbital magnetization (OM), which incorporates contributions from both bulk and surface Wannier functions~\cite{thonhauser2005orbital,thonhauser2011theory,ceresoli2006orbital,hanke2016role,marzari1997maximally,souza2001maximally,marzari2012maximally,shi2007quantum,go2024first}. Importantly, Wannier interpolation applied to OM has so far been the most numerically stable scheme for evaluating OM in transition metals~\cite{lopez2012wannier}.

Here, we show that the framework of Wannier functions enables the quantification of the orbital current operator within Wannier space, facilitating the computation of various orbital transport quantities. Following the spirit of the Wannier interpolation approach, we extend the modern theory of OM to derive a fully gauge-covariant expression for the orbital current operator in the basis of Bloch-like Wannier functions, showing that it can be entirely quantified in terms of Berry-type quantities~\cite{lopez2012wannier,thonhauser2005orbital,thonhauser2011theory}. We then apply our theory to the evaluation of the OHE in selected material systems~\cite{pizzi2020wannier90,marzari1997maximally,marzari2012maximally,souza2001maximally,fleurWeb}. Crucially, the inherent non-locality of the Wannier-based OAM operator gives rise to an additional non-Kubo contribution when calculating quantities like the OHC that is absent from previous treatments \cite{pezo2022orbital,bhowal2021orbital}. We thereby provide the first detailed account of the anatomy and relative importance of local and itinerant contributions to the OHE across transition metals and a range of other selected materials, contributing to the identification of next-generation orbitronic devices.

\section*{Results}

\subsection*{OAM operator}

\quad According to the modern theory of orbital magnetization, OM is separated into local-circulation (LC) and itinerant-circulation (IC) contributions~\cite{lopez2012wannier,hanke2016role,thonhauser2005orbital,thonhauser2011theory,ceresoli2006orbital}. LC originates from the self-rotation of Wannier functions around their respective centers, whereas IC arises from the motion of Wannier function centers~\cite{lopez2012wannier,ceresoli2006orbital,thonhauser2005orbital,thonhauser2011theory}. The modern OM is also directly dependent on the chemical potential within the gap of insulators with a nonzero Chern number~\cite{lopez2012wannier,thonhauser2005orbital,thonhauser2011theory,ceresoli2006orbital,shi2007quantum,hanke2016role}.

Our derivation starts from the real-space matrix element of the OAM operator between Wannier functions $\vert n\textbf{R} \rangle$ centered at lattice site $\textbf{R}$, which we split as $\langle n\textbf{R} \vert \hat{\textbf{L}} \vert m\textbf{R} \rangle = \langle n\textbf{R} \vert \left( \hat{\textbf{r}} - \textbf{R} \right) \times \hat{\textbf{p}} \vert m\textbf{R} \rangle + \textbf{R} \times \langle n\textbf{R} \vert \hat{\textbf{p}} \vert m\textbf{R} \rangle$, i.e.\ into a term local to the Wannier center and a term proportional to the lattice vector itself. Following Thonhauser \textit{et al.}~\cite{thonhauser2005orbital}, we divide the crystal into a bulk and a surface region, with the boundary placed deep enough that the surface region is still faithfully described by bulk-like Wannier functions. In the thermodynamic limit, summing over all sites shows that the local term receives contributions exclusively from the bulk, while the term proportional to $\textbf{R}$ receives contributions exclusively from the surface~\cite{thonhauser2005orbital,thonhauser2011theory,ceresoli2006orbital}; this is the real-space origin of the LC/IC split. Crucially, because the surface contribution can still be expressed using bulk Wannier functions, the resulting expressions are independent of the specific surface details of the solid, eliminating the need for special surface Wannier functions and allowing the OAM operator to be evaluated entirely from bulk quantities~\cite{thonhauser2005orbital,ceresoli2006orbital,thonhauser2011theory}. This can be utilized to derive an expression for the general matrix elements of the OAM operator by treating the position operator in these regions with care, an aspect not considered in previous approaches~\cite{pezo2022orbital,busch2023orbital}.

Our method requires switching between two gauges: the Wannier gauge and the Hamiltonian gauge~\cite{thonhauser2005orbital,thonhauser2011theory,ceresoli2006orbital,lopez2012wannier,marzari1997maximally,souza2001maximally}. The Bloch-like functions in the Wannier gauge are derived by transforming the Wannier states $\vert n\textbf{R} \rangle$ into reciprocal space, $\vert u_{n\textbf{k}}^{\textrm{W}} \rangle = \frac{1}{\sqrt{N}} \sum_{\textbf{R}} e^{i\textbf{k} \cdot \left( \textbf{R} - \textbf{r} \right)} \vert n\textbf{R} \rangle$, whereas the Hamiltonian gauge serves as the eigenspace of the Bloch Hamiltonian $\hat{H}_{\textbf{k}} =  e^{-i \textbf{k} \cdot \textbf{r}} \hat{\mathcal{H}} e^{i \textbf{k} \cdot \textbf{r}}$ projected onto the $M$-dimensional inner space, and can be obtained through the transformation $\vert u_{n\textbf{k}}^{\textrm{H}} \rangle = \sum_{m}^{M} \vert u_{m\textbf{k}}^{\textrm{W}} \rangle U_{mn}$~\cite{lopez2012wannier,go2024first}. With "inner space", we refer to the space spanned by the Wannier functions, which only becomes a complete basis set when combining it with the "outer space", not included in the Wannier representation \cite{lopez2012wannier}. As we show in the Supplementary Material, the expression for the general matrix elements of the OAM operator at any $k$-point in the Hamiltonian gauge $\textbf{L}_{nm}^{\textrm{H}}=\langle  u_{n\textbf{k}}^{\textrm{H}} \vert \hat{\textbf{L}} \vert u_{m\textbf{k}}^{\textrm{H}} \rangle$, reads:
\begin{equation} \label{AngularMomMethod2}
\textbf{L}_{nm}^{\textrm{H}}=
 \frac{ie}{2\mu_{B}} \langle \partial_{\textbf{k}} u_{n\textbf{k}}^{\textrm{H}} \vert \times \left( \hat{H}_{\textbf{k}} + \frac{\varepsilon_{n\textbf{k}} + \varepsilon_{m\textbf{k}} }{2} - 2 \varepsilon_{F} \right) \vert \partial_{\textbf{k}}u_{m\textbf{k}}^{\textrm{H}} \rangle .
\end{equation}
Here, $\varepsilon_{n\textbf{k}}$ represents the eigenenergy associated with the Bloch Hamiltonian eigenstate $\vert u^{\rm H}_{n\textbf{k}} \rangle$\footnote{These energies are generally not equivalent to the ab-initio eigenenergies due to the Wannierization process~\cite{lopez2012wannier,pizzi2020wannier90}. While the occupied and some low-lying empty energy bands are conserved by the disentanglement, higher-lying bands will undergo changes as a result~\cite{lopez2012wannier,marzari2012maximally,souza2001maximally,pizzi2020wannier90}.}. However, this expression is neither our final result nor the proper way of treating OAM in solids. We will resolve this in the following section, by introducing a covariant derivative.

Following the approach by Lopez and co-workers, the OAM expression can be rewritten by utilizing Berry-type quantities~\cite{lopez2012wannier}. In the Wannier gauge, these quantities are
$\mathbb{A}_{nm} = i \langle u_{n\textbf{k}}^{\textrm{W}} \vert \partial_{\textbf{k}} u_{m\textbf{k}}^{\textrm{W}} \rangle$, $
\mathbb{B}_{nm} = i \langle u_{n\textbf{k}}^{\textrm{W}} \vert \hat{H}_{\textbf{k}} \vert \partial_{\textbf{k}} u_{m\textbf{k}}^{\textrm{W}} \rangle$,
$\mathbb{\Omega}_{nm} = i \langle \partial_{\textbf{k}} u_{n\textbf{k}}^{\textrm{W}} \vert \times \vert \partial_{\textbf{k}} u_{m\textbf{k}}^{\textrm{W}} \rangle$, and
$\mathbb{\Lambda}_{nm} = i \langle \partial_{\textbf{k}} u_{n\textbf{k}}^{\textrm{W}} \vert \times \hat{H}_{\textbf{k}} \vert \partial_{\textbf{k}} u_{m\textbf{k}}^{\textrm{W}} \rangle$,
which, when transformed into Hamiltonian gauge, become~\cite{lopez2012wannier}
\begin{eqnarray}
A_{nm} &= \overline{\mathbb{A}}_{nm} + \overline{J}_{nm} , \\
B_{nm} &= \overline{\mathbb{B}}_{nm} + \sum_{m'} \overline{\mathbb{H}}_{nm'} \overline{J}_{m'm} , \\
\Omega_{nm} &= \left(\overline{\mathbb{\Omega}} + i\overline{J} \times \overline{\mathbb{A}} + i\overline{\mathbb{A}} \times \overline{J} + i\overline{J} \times \overline{J} \right) _{nm} , \\
\Lambda_{nm} &= \left( \overline{\mathbb{\Lambda}} + i\overline{J} \times \overline{\mathbb{B}} + i\overline{\mathbb{B}}^{\dagger} \times \overline{J} + i\overline{J} \times \overline{\mathbb{H}} \overline{J} \right) _{nm} .
\end{eqnarray}
Here, we introduce the notation $\overline{X} = U^{\dagger} X U$ for an arbitrary matrix $X$ in Wannier gauge and $\mathbb{H}_{nm} = \langle u_{n\textbf{k}}^{\textrm{W}} \vert \hat{H}_{\textbf{k}} \vert u_{m\textbf{k}}^{\textrm{W}} \rangle$. Transforming the Berry-type matrices from Wannier into Hamiltonian gauge gives rise to contributions containing the Hermitian matrix
$\overline{J} = i U^{\dagger}\partial_{\textbf{k}} U$, which contains poles at band crossings and can lead to numerical instabilities~\cite{lopez2012wannier}. Unlike the Berry-type matrices, $\overline{J}$ can be defined directly in Hamiltonian gauge as~\cite{lopez2012wannier}
\begin{eqnarray}
\overline{J}_{nm} = i \left[ U^{\dagger} \partial_{\textbf{k}} U \right]_{nm} = \begin{cases} \frac{i\left[ U^{\dagger} \left(\partial_{\textbf{k}}\mathbb{H}\right) U \right]_{nm}}{\varepsilon_{m\textbf{k}} - \varepsilon_{n\textbf{k}}} &\textrm{if } n \neq m \\ 0 &\textrm{if } n=m .
\end{cases}
\end{eqnarray}
Using all of these quantities, we are able to reformulate the matrix elements of the OAM operator in Hamiltonian gauge as
\begin{equation}
\begin{split}
&
\textbf{L}_{nm}^{\textrm{H}} =  \frac{e}{2\mu_{B}} \Big[ \overline{\mathbb{\Lambda}} + i\overline{J} \times \overline{\mathbb{B}} + i\overline{\mathbb{B}}^{\dagger} \times \overline{J} + i\overline{J} \times \overline{\mathbb{H}} \overline{J} \\& + \frac{1}{2}\left\{\overline{\mathbb{H}} - 2\varepsilon_{F}, \overline{\mathbb{\Omega}} + i\overline{J} \times \overline{\mathbb{A}} + i\overline{\mathbb{A}} \times \overline{J} + i\overline{J} \times \overline{J} \right\} \Big]_{nm} .
\end{split}
\end{equation}

Note that this expression is not covariant and will lead to uncontrolled terms when evaluated. We therefore introduce a covariant derivative in the following section that eliminates those problems by replacing the partial derivative in Eq.~(\ref{AngularMomMethod2}).

\subsection*{Covariant derivative and modern OAM operator}

\quad We now introduce a covariant derivative to deal with the gauge dependence systematically and obtain an expression for the OAM operator shown in Eq.~(\ref{AngularMomMethod2}) that is covariant under any gauge transformation. At the same time, this prevents the occurrence of numerical instabilities in the evaluation of $\overline{J}$ when band crossings occur and controls the emerging poles in the OAM operator. In the context of OM, the covariant derivative $D_{\textbf{k}}$ is usually defined by projecting the conventional derivative of occupied states onto the unoccupied space $\vert D_{\textbf{k}} u_{n\textbf{k}} \rangle = \hat{Q}_{\textbf{k}} \vert \partial_{\textbf{k}}u_{n\textbf{k}} \rangle$, with $\hat{Q}_{\textbf{k}} = 1 - \hat{P}_{\textbf{k}} = 1 - \sum_{n} \vert u_{n\textbf{k}}^{\textrm{H}} \rangle f_{n\textbf{k}}^{\textrm{H}} \langle u_{n\textbf{k}}^{\textrm{H}} \vert $, with $f_{n\textbf{k}}^{\textrm{H}}$ being the Fermi-occupation function in Hamiltonian gauge~\cite{wang2006ab,ceresoli2006orbital,lopez2012wannier}. However, because we also consider the unoccupied matrix elements of the OAM operator, we must extend this definition so that it also projects the derived inner unoccupied states ($\hat{Q}_{\textrm{in},\textbf{k}}$) onto their orthogonal space. The orthogonal space of the inner unoccupied space includes not only all occupied states ($\hat{P}_{\textbf{k}}$) but also all outer states ($\hat{\mathbb{Q}}_{\textbf{k}}$), which are not spanned by the Wannier basis~\cite{lopez2012wannier,lee2026anatomy}. The modified covariant derivative can be expressed as follows:
\begin{equation} \label{CovDerOccUnocc}
\begin{split}
\vert D_{\textbf{k}} u_{n\textbf{k}}^{\textrm{H}} \rangle & = \hat{Q}_{\textbf{k}} \vert \partial_{\textbf{k}}u_{n\textbf{k}}^{\textrm{H}} \rangle f_{n}^{\textrm{H}} + \left( \hat{P}_{\textbf{k}} + \hat{\mathbb{Q}}_{\textbf{k}} \right) \vert \partial_{\textbf{k}}u_{n\textbf{k}}^{\textrm{H}} \rangle g_{n}^{\textrm{H}} \\ & = \vert \partial_{\textbf{k}} u_{n\textbf{k}}^{\textrm{H}} \rangle + i \sum_{m} \vert u_{m\textbf{k}}^{\textrm{H}} \rangle \left( f_{m}^{\textrm{H}} A_{mn} f_{n}^{\textrm{H}} + g_{m}^{\textrm{H}} A_{mn} g_{n}^{\textrm{H}} \right) .
\end{split}
\end{equation}

This definition possesses a neat symmetry between the occupation functions of the occupied ($f$) and inner unoccupied ($g = 1 - f$) spaces. Replacing the regular derivative in Eq.~(\ref{AngularMomMethod2}) by the covariant one yields what we refer to as the \textit{modern OAM operator}. As we show in the Supplementary Material, the modern OAM operator produces the modern expression for the ground-state OM, obtained as a trace over all occupied states of $\textbf{L}_{nm}^{\textrm{H}}$. There, we also demonstrate how the final expression of the modern OAM operator, where the $k$-derivatives in Eq.~(\ref{AngularMomMethod2}) are replaced with covariant derivatives

\begin{equation} \label{AngularMomMethod3}
\textbf{L}_{nm}^{\textrm{H}}=
 \frac{ie}{2\mu_{B}} \langle D_{\textbf{k}} u_{n\textbf{k}}^{\textrm{H}} \vert \times \left( \hat{H}_{\textbf{k}} + \frac{\varepsilon_{n\textbf{k}} + \varepsilon_{m\textbf{k}} }{2} - 2 \varepsilon_{F} \right) \vert D_{\textbf{k}}u_{m\textbf{k}}^{\textrm{H}} \rangle ,
\end{equation}
can be expressed in Hamiltonian gauge using Wannier Berry-type quantities. Additionally, we demonstrate that this expression can be reformulated to consist entirely of projectors and the Hamiltonian, which directly proves the gauge covariance of the matrix elements produced by the modern OAM operator~\cite{wang2006ab}. Schematically, $\hat{\textbf{L}} \propto \left( \partial_{\textbf{k}} \hat{P}_{\textbf{k}} \right) \hat{Q}_{\textbf{k}} \times \hat{H}_{\textbf{k}} \hat{Q}_{\textbf{k}} \left( \partial_{\textbf{k}} \hat{P}_{\textbf{k}} \right) + \ldots$, where $\hat{P}_{\textbf{k}}$ and $\hat{Q}_{\textbf{k}}$ project onto the occupied and unoccupied subspaces, respectively; since projectors and the Hamiltonian are manifestly gauge-independent objects, this form makes the gauge covariance of the modern OAM operator explicit (full expression in the Supplementary Material). In addition to achieving gauge covariance, another benefit of introducing the covariant derivative now becomes clear: according to Eq.~({\color{blue}11}) of the Supplementary Material, the $J$-matrix now appears only enveloped by the occupation functions $f$ and $g$, which suppresses the emergence of poles when computing the corresponding terms, resulting in higher numerical stability. Note that poles can still emerge due to the broadening of the occupation functions at finite temperature, or if a band crossing occurs exactly at the Fermi level. To account for both cases, we add a small constant $i\eta$ to the denominator of $\overline{J}$, as detailed in Eq.~({\color{blue}12}) of the Supplementary Material. These crucial properties are absent from previous expressions of the OAM operator that attempt to go beyond the limits of atomic approximations, and their absence can substantially alter the obtained results~\cite{pezo2022orbital,busch2023orbital}. This highlights the importance of treating the position operator and the space projections rigorously.

The derived expression for the modern OAM operator holds exactly in the zero-temperature limit; however, it remains a valid approximation at low temperatures. As before, this expression can be separated into its LC and IC part ($\hat{\textbf{L}} = \hat{\textbf{L}}^{\textrm{LC}} + \hat{\textbf{L}}^{\textrm{IC}}$), where $\hat{\textbf{L}}^{\textrm{IC}}$ becomes directly proportional to the difference between the band energy and the Fermi-energy when evaluated in Hamiltonian gauge, while $\textbf{L}^{\textrm{LC}}$ collects the remaining terms. Tracing $f\hat{\textbf{L}}$ or only its LC/IC contribution reproduces the ground-state OM or its respective LC/IC counterpart of Ref.~\cite{lopez2012wannier} exactly, confirming the consistency of our operator with the established modern theory of orbital magnetization. 

\subsection*{Dynamic gauge and orbital Hall conductivity}

\quad For computing the OHC we introduce a special gauge, which we refer to as the dynamic gauge, $\vert u_{n\textbf{k}}^{\textrm{D}} \rangle = \sum_{m} \vert u_{m\textbf{k}}^{\textrm{H}} \rangle V_{mn}\left(t\right)$, where $V\left(t\right)$ captures the effect of a time-dependent perturbation. For evaluating the orbital current response, we perturb the system by an electric field $\hat{H}_{\textbf{k}}^{(1)} = e \textbf{E} \cdot \hat{\textbf{r}}$. The electric field not only modifies the states but also changes the occupation functions, with both contributing to the OHC. The latter contribution, which cannot be quantified using the dynamic gauge and can instead be treated by, for example, applying the relaxation time approximation~\cite{atwal2002relaxation}, is neglected in the following. In first order, the transformation matrix $V$ can be approximated as $V = 1 + \delta V$. For the considered small time-independent electric perturbation, this yields
\begin{eqnarray} \label{EPertubation}
\delta V_{nm} = ie\hbar\frac{\langle u_{n\textbf{k}}^{\textrm{H}}\vert \textbf{E} \cdot \hat{\textbf{v}}\vert u_{m\textbf{k}}^{\textrm{H}} \rangle}{\left( \varepsilon_{n\textbf{k}} - \varepsilon_{m\textbf{k}}\right)^{2} + \eta^{2}} ,
\end{eqnarray}
with $\delta V_{nm} = 0$ for $n = m$ and the velocity operator in Hamiltonian gauge~\cite{go2024first}

\begin{eqnarray} \label{VelocityOp}
\hat{\textbf{v}} = \frac{1}{\hbar} \sum_{mm',\textbf{k}} \vert u^{\textrm{H}}_{m\textbf{k}}\rangle \Big[ \delta_{mm'} \partial_{\textbf{k}} \varepsilon_{m\textbf{k}} + i \left( \varepsilon_{m\textbf{k}} - \varepsilon_{m'\textbf{k}} \right) A_{mm'} \Big] \langle u^{\textrm{H}}_{m'\textbf{k}} \vert.
\end{eqnarray}
The small broadening $\eta$ is added to avoid poles and ensure convergence of the time integral of $V$. When transforming between Wannier and dynamic gauges, the transformation matrix and the matrix $\overline{J}$ become $U' = U + U\delta V$ and $\overline{J}_{\alpha}' = \overline{J}_{\alpha} + \delta V^{\dagger} \overline{J}_{\alpha} + \overline{J}_{\alpha} \delta V + \overline{\tilde{J}}$ with $\overline{\tilde{J}} = i \partial_{\alpha} \delta V$.
The OHC
\begin{eqnarray} \label{OHCfromj}
\sigma_{\alpha\beta}^{L_{\gamma}} = \int_{\textrm{BZ}} \frac{d^{d}\textbf{k}}{\left( 2\pi \right)^{d}} \frac{ \partial \langle j_{\alpha}^{L_{\gamma}} \rangle^{\textrm{D}}}{\partial E_{\beta}} \Bigg|_{\textbf{E} \rightarrow \textbf{0}} = \int_{\textrm{BZ}} \frac{d^{d}\textbf{k}}{\left( 2\pi \right)^{d}} \frac{\partial \langle\delta j_{\alpha}^{L_{\gamma}} \rangle^{\textrm{D}}}{\partial E_{\beta}} ,
\end{eqnarray}
is proportional to the orbital current $j_{\alpha}^{L_{\gamma}}$ generated by the electric perturbation~\cite{salemi2022first,busch2023orbital}. Starting from the orbital current operator $j_{\alpha}^{L_{\beta}} = \frac{1}{2} \left\{ \hat{v}_{\alpha}, \hat{L}_{\beta} \right\}$, we can identify two contributions to the change in orbital current in first-order perturbation theory: $\langle \delta j_{\alpha}^{L_{\beta}} \rangle^{\textrm{D}} = \langle \delta j_{\alpha}^{L_{\beta}} \rangle^{\textrm{D}}_{\textrm{I}} + \langle \delta j_{\alpha}^{L_{\beta}} \rangle^{\textrm{D}}_{\textrm{II}}$. The first can be interpreted as the change in velocity, or the rate at which the OAM is transported, $\langle \delta j_{\alpha}^{L_{\beta}} \rangle^{\textrm{D}}_{\textrm{I}} = \frac{1}{2} \sum_{nm} f_{n}^{\textrm{H}} \left[ \delta v_{\alpha,nm}^{\textrm{D}} L_{\beta,mn}^{\textrm{H}} + L_{\beta,nm}^{\textrm{H}} \delta v_{\alpha,mn}^{\textrm{D}} \right]$, and the second as a change in the magnitude of the transported OAM, $\langle \delta j_{\alpha}^{L_{\beta}} \rangle^{\textrm{D}}_{\textrm{II}} = \frac{1}{2} \sum_{nm} f_{n}^{\textrm{H}} \left[ v_{\alpha,nm}^{\textrm{H}} \delta L_{\beta,mn}^{\textrm{D}} + \delta L_{\beta,nm}^{\textrm{D}} v_{\alpha,mn}^{\textrm{H}} \right]$. 

While the velocity operator is a local operator whose first-order change can therefore be evaluated simply as $\delta v_{\alpha}^{\textrm{D}} = \delta V^{\dagger} v_{\alpha}^{\textrm{H}} + v_{\alpha}^{\textrm{H}} \delta V$, the same does not hold for the modern OAM operator. This is due to the non-locality of the modern OAM operator encoded in the $J$-matrices, which yields an additional contribution when brought into dynamic gauge:
\begin{eqnarray} \label{dLD}
\delta L^{\textrm{D}}_{\alpha} = \delta V^{\dagger} L_{\alpha}^{\textrm{H}} + L_{\alpha}^{\textrm{H}} \delta V + \delta \tilde{L}_{\alpha},
\end{eqnarray}
where the additional contribution appears as $\delta \tilde{L}_{\alpha}$. The term $\delta \tilde{L}_{\alpha}$ is obtained by replacing all of the $J$-matrices in the OAM operator by $\tilde{J}$ in a way that makes the expression proportional to $\tilde{J}$, where, for $n \neq m$,
\begin{equation} \label{Jtilde}
\overline{\tilde{J}}_{\alpha,nm} = -e\hbar\,\partial_{\alpha} \frac{\langle u_{n\textbf{k}}^{\textrm{H}}\vert \textbf{E} \cdot \hat{\textbf{v}}\vert u_{m\textbf{k}}^{\textrm{H}} \rangle}{\left( \varepsilon_{n\textbf{k}} - \varepsilon_{m\textbf{k}}\right)^{2} + \eta^{2}} ,
\end{equation}
and $\overline{\tilde{J}}_{\alpha,nm}=0$ for $n=m$ holds. Similar to $\overline{J}$, $\overline{\tilde{J}}$ carries non-local contributions important for a full description of the OHC and possibly other orbital quantities. An OAM operator restricted to atom-centered, local matrix elements would yield $\overline{\tilde{J}} = 0$ and no such term would appear. The full expression for $\delta\tilde{L}_{\alpha}$ in terms of $\overline{\tilde{J}}$ and the Berry-type quantities can be found in the Supplementary Material. 

We are now able to calculate the first-order orbital current generated by the electric perturbation:

\begin{equation} \label{jFull}
\begin{split}
\langle \delta j_{\alpha}^{L_{\beta}} \rangle^{\textrm{D}} = & \sum_{nm} f_{n}^{\textrm{H}} \Big[ \delta V^{\dagger}_{nm} j_{\alpha,mn}^{L_{\beta}} + j_{\alpha,nm}^{L_{\beta}} \delta V_{mn} + \frac{1}{2} \left( v_{\alpha,nm}^{\textrm{H}} \delta \tilde{L}_{\beta,mn} + \delta \tilde{L}_{\beta,nm} v_{\alpha,mn}^{\textrm{H}} \right) \Big] \\ = & \langle \delta j_{\alpha}^{L_{\beta}} \rangle^{\textrm{D}}_{\textrm{Kubo}} + \langle \delta j_{\alpha}^{L_{\beta}} \rangle^{\textrm{D}}_{\textrm{Non-Kubo}},
\end{split}
\end{equation}
which, based on Eq.~(\ref{OHCfromj}), can be rewritten for the OHC as a sum of the well-known Kubo formula and an additional non-Kubo term containing $\delta\tilde{L}_{\alpha}$. This additional term does not appear in previous approaches due to the absence of the non-local OAM contributions~\cite{pezo2022orbital,busch2023orbital}. Similar non-Kubo terms giving large contributions could also arise for other quantities containing the OAM operator. We therefore stress that the integration of the modern OAM operator into other theories cannot be reduced to a simple replacement of a previously used OAM operator with the modern one in the corresponding expressions.

By performing first-principles-based Wannier interpolation calculations (see details in the Methods section), we obtain estimates of the OHC in various specific material systems using the developed formalism~\cite{fleurCode,fleurWeb,pizzi2020wannier90}, with results presented in Table~\ref{Table1}.

\begin{figure}[htbp]
\includegraphics[angle=0, width=0.5\textwidth]{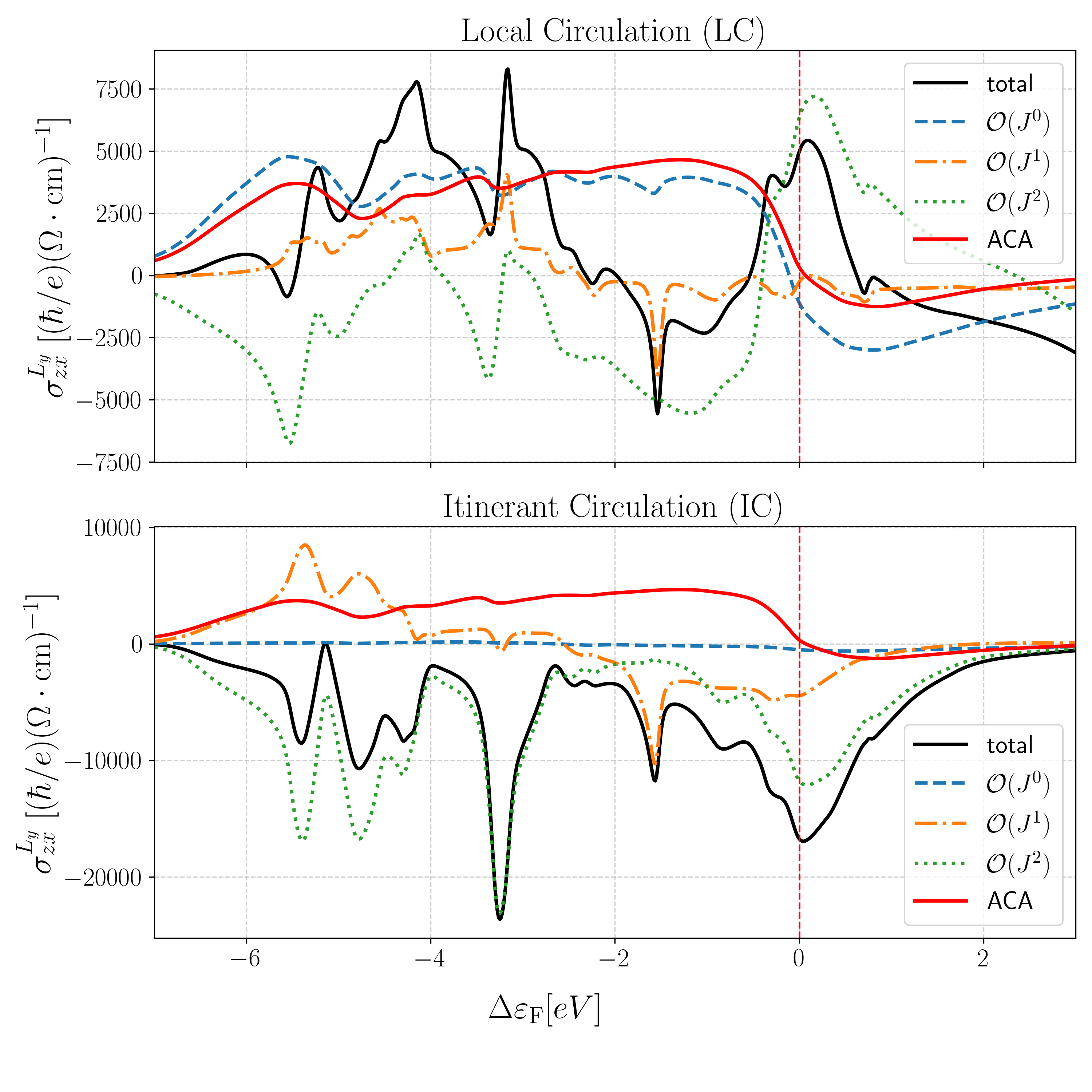}
\centering
\caption{
\label{OHCPlot}
Modern OHC $\sigma_{zx}^{L_{y}}$ in fcc Platinum plotted as a function of the Fermi energy, separated into its contributions emerging from LC and IC, respectively, shown in orders of $J$ and compared to the ACA.}
\end{figure}

We start by considering the case of fcc Platinum in detail. The black solid lines in the upper and lower panels in Fig. 1 show, respectively, the calculated LC and IC contributions to the OHC in the modern approach. Each contribution is decomposed further into contributions $\mathcal{O}\left(J^{0}\right)$, $\mathcal{O}\left(J^{1}\right)$, $\mathcal{O}\left(J^{2}\right)$, which do not contain $J$, do contain $J$ in the linear order, and do contain $J$ in the quadratic order, respectively. The OHC obtained with the ACA (red solid line) is presented both in the upper and lower panels to compare it with the contributions from the modern approach. As shown in Fig.~\ref{OHCPlot}, the $\mathcal{O}\left(J^{0}\right)$ LC contribution in the modern approach agrees excellently with the OHC obtained by the ACA in the entire energy window examined. Thus, the $\mathcal{O}\left(J^{0}\right)$ LC contribution may be interpreted as the modern approach description of the ACA. Note that other contributions in the modern approach are not negligible compared to the $\mathcal{O}\left(J^{0}\right)$ LC contribution. Thus, there is a drastic difference between the OHC values computed within the ACA and those obtained using the total modern OAM operator. This concerns both the LC and IC contributions, where the LC contribution also carries non-local terms in the form of the $J$-matrices that drive this difference.

In contrast to the ACA and $\mathcal{O}\left( J^{0} \right)$ OHC, all other contributions display a highly non-trivial dependence on band filling, suggesting that the non-local contributions to the OHC are very sensitive to the electronic structure details of a specific material. This is in direct contradiction to the current common understanding of the OHE~\cite{mankovsky2024spin,go2018intrinsic}. Another observation is that the LC and IC contributions to the OHC are of the same order of magnitude but opposite sign at the Fermi energy, which ultimately results in an OHC with opposite sign at the true Fermi energy of Pt compared to the ACA result. This suggests that the IC contribution can dominate the OHC in certain systems and dictate its sign, highlighting the importance of accounting for all non-local OAM contributions when assessing non-equilibrium orbital effects. This stands in contrast to the equilibrium OM, for which LC contributions are understood to dominate in systems where electron orbitals are well localized~\cite{lopez2012wannier,ceresoli2010first,lee2026anatomy}.

\begin{table}[htbp]
\centering
\caption{Modern OHC and its different contributions, and ACA OHC ($\sigma_{zx}^{L_{y}}$) at the estimated true Fermi energy in units of $10^3\cdot\frac{\hbar}{e}\left(\Omega \cdot \textrm{cm} \right)^{-1}$. We consider fcc Pt, bcc W, ferromagnetic bcc Fe, bcc V, antiferromagnetic bcc Cr, $\alpha$-Ge (diamond structure), fcc Cu, hcp Ti, and monolayer 2H-MoS$_2$. For monolayer MoS$_2$ $\sigma_{yx}^{L_{z}}$ is considered instead.}
\arrayrulecolor{black}
\setlength{\arrayrulewidth}{0.3pt}
\renewcommand{\arraystretch}{1.2}
\begin{tabular}{@{}c|ccccccccc@{}}
\hhline{----------}
\rowcolor{gray!15}
\textbf{} & \textbf{Total} & \textbf{LC} & \textbf{IC} & \textbf{Kubo} & \textbf{Non-Kubo} & $\mathcal{O}\left( J^{0} \right)$ & $\mathcal{O}\left( J^{1} \right)$ & $\mathcal{O}\left( J^{2} \right)$ & \textbf{ACA} \\
\hhline{-|---------}
\rowcolor{white}
\textbf{fcc Pt} & $-$11.75 & \phantom{$-$}4.98 & $-$16.73 & $-$15.84 & \phantom{$-$}4.09 & $-$1.61 & $-$4.72 & $-$5.42 & \phantom{$-$}0.33 \\
\rowcolor{gray!10}
\textbf{bcc W} & \phantom{$-$}4.05 & \phantom{$-$}7.62 & $-$3.57 & \phantom{$-$}4.35 & $-$0.30 & \phantom{$-$}6.04 & \phantom{$-$}4.41 & $-$6.40 & \phantom{$-$}4.62 \\
\rowcolor{white}
\textbf{bcc Fe} & $-$15.56 & $-$0.21 & $-$15.35 & $-$15.30 & $-$0.26 & \phantom{$-$}2.31 & \phantom{$-$}1.36 & $-$19.23 & \phantom{$-$}2.54 \\
\rowcolor{gray!10}
\textbf{bcc V} & \phantom{$-$}2.40 & \phantom{$-$}3.19 & $-$0.79 & \phantom{$-$}2.33 & \phantom{$-$}0.07 & \phantom{$-$}4.61 & \phantom{$-$}2.86 & $-$5.07 & \phantom{$-$}4.30 \\
\rowcolor{white}
\textbf{bcc Cr} & $-$7.40 & \phantom{$-$}2.34 & $-$9.75 & $-$6.53 & $-$0.87 & \phantom{$-$}4.58 & \phantom{$-$}1.86 & $-$13.84 & \phantom{$-$}4.08 \\
\rowcolor{gray!10}
\textbf{$\alpha$-Ge} & $-$2.23 & \phantom{$-$}1.12 & $-$3.35 & $-$1.59 & $-$0.64 & $-$0.28 & $-$1.00 & $-$0.95 & $-$0.55 \\
\rowcolor{white}
\textbf{fcc Cu} & $-$2.22 & $-$1.65 & $-$0.57 & $-$1.92 & $-$0.30 & $-$1.58 & $-$0.24 & $-$0.40 & $-$0.67 \\
\rowcolor{gray!10}
\textbf{hcp Ti} & $-$3.30 & \phantom{$-$}4.03 & $-$7.33 & $-$3.50 & \phantom{$-$}0.20 & \phantom{$-$}4.50 & \phantom{$-$}6.50 & $-$14.30 & \phantom{$-$}4.40 \\
\rowcolor{white}
\textbf{MoS$_2$} & $-$0.38 & $-$1.82 & \phantom{$-$}1.44 & $-$0.74 & \phantom{$-$}0.36 & $-$1.11 & \phantom{$-$}0.08 & \phantom{$-$}0.65 & $-$0.93 \\
\hhline{----------}
\end{tabular}
\label{Table1}
\end{table}

\begin{figure}[htbp]
    \centering
    \setlength{\abovecaptionskip}{0pt}
    \setlength{\belowcaptionskip}{0pt}

    \begin{subfigure}{\linewidth}
        \centering
        \includegraphics[width=0.5\linewidth]{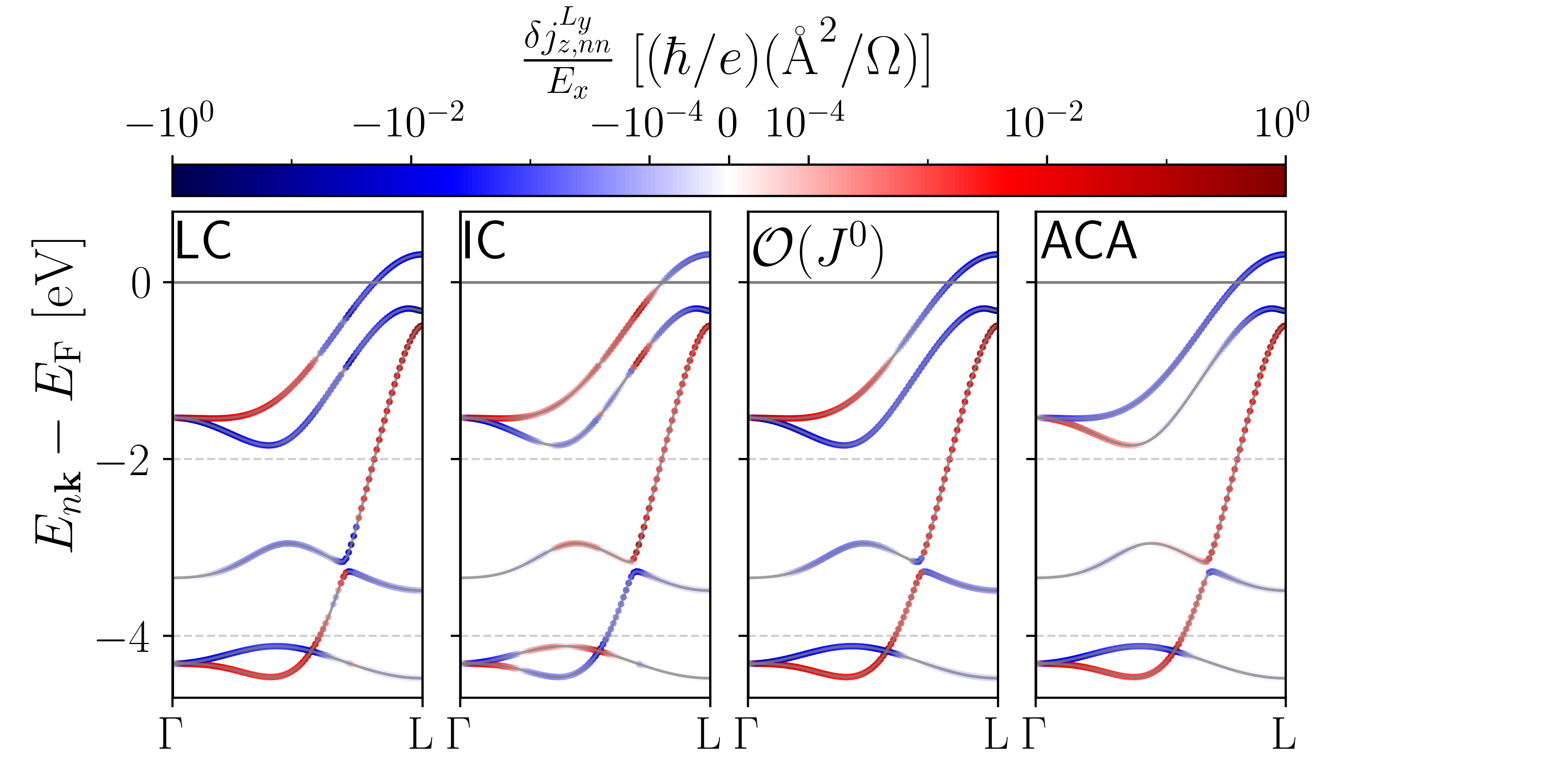}
    \end{subfigure}

   \vspace{-2pt}

    \begin{subfigure}{\linewidth}
        \centering
        \includegraphics[width=0.5\linewidth]{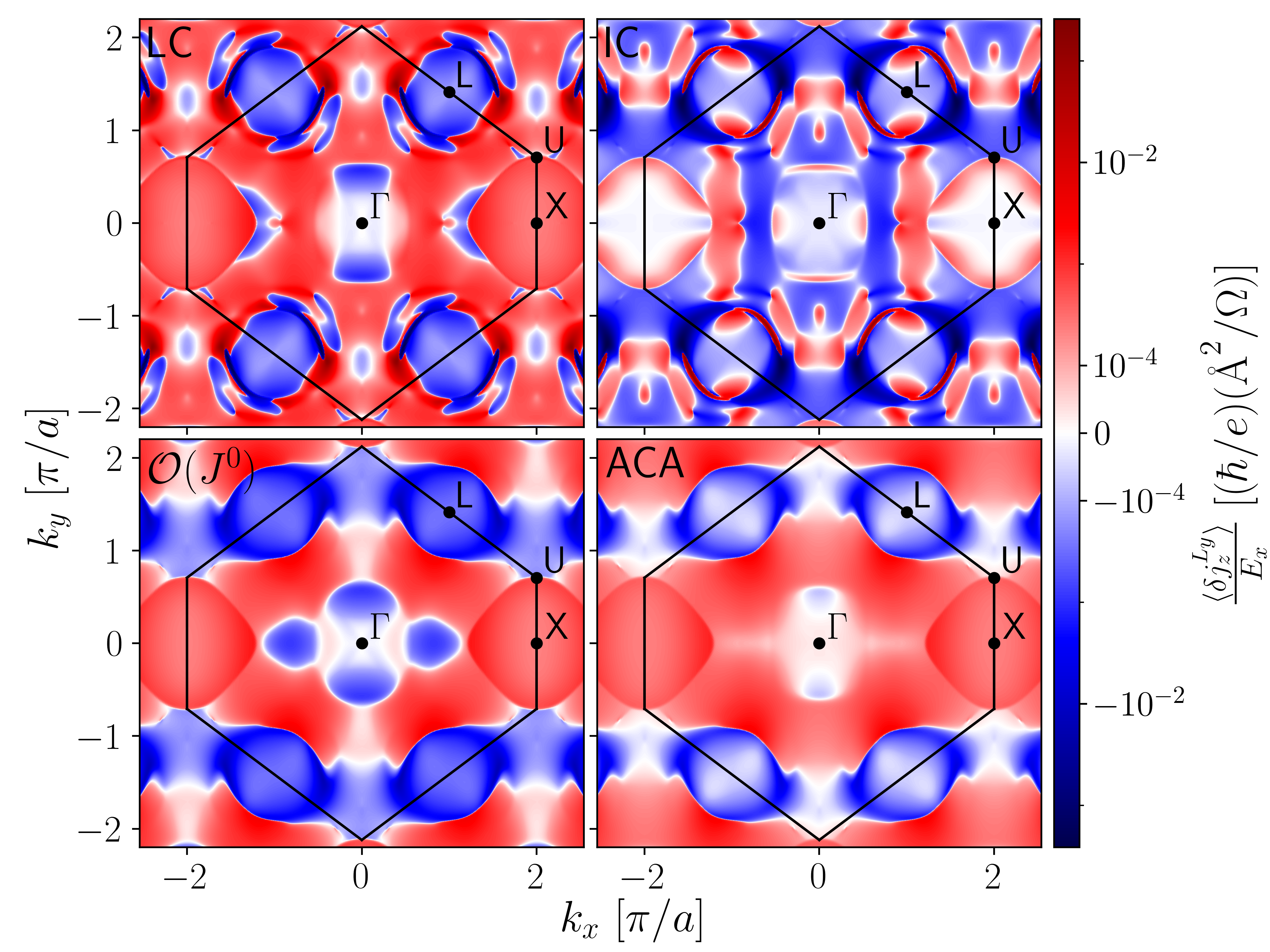}
    \end{subfigure}
    \centering
    \caption{Band- and $k$-resolved LC, IC, $\mathcal{O}\left(J^{0}\right)$, and ACA contributions to the OHC between the high-symmetry points $\Gamma$ and L for fcc Pt (upper panels: a,b,c,d), and the corresponding $k$-resolved OHCs for a cut through the first Brillouin zone (lower panels: e,f,g,h).}

    \label{fig2}
\end{figure}

The strong impact of the itinerant OHC on the overall effect is apparent from Table~\ref{Table1}. For many of the considered materials, the IC contribution is larger in magnitude and of opposite sign to the LC contribution. The only exceptions that share the same sign are ferromagnetic bcc Fe and fcc Cu. Comparison of the $\mathcal{O}\left(J^{0}\right)$ contribution with the ACA OHC confirms our findings for fcc Pt: both quantities share a very similar behaviour. Except for fcc Pt, they are also consistently similar in magnitude and carry the same sign across all materials. We confirm this trend further by examining the band-filling dependence of the two contributions in all considered materials, and observe that it holds relatively well throughout $k$-space, as demonstrated for example by the distribution of ACA and $\mathcal{O}\left(J^{0}\right)$ OHC along the high-symmetry lines and at the Fermi energy of fcc Pt projected onto the $(k_x,k_y,0)$-plane, shown in Fig.~\ref{fig2}(c,d) and (g,h), respectively. Another noteworthy property of the modern OHC is that the $\mathcal{O}\left(J^{2}\right)$ part appears to dominate in most systems, most prominently in bcc Fe, bcc Cr, and hcp Ti. The opposite can be observed for the non-Kubo part of the OHC, which appears to be much smaller than its Kubo counterpart for most of the studied systems, with the exception of fcc Pt, $\alpha$-Ge, and monolayer MoS$_{2}$. Most strikingly, however, the sign and magnitude of the total modern OHC can differ substantially from the conventional ACA values, which are widely used to confirm the orbital nature of observed phenomena in various experiments~\cite{ando2025orbitronics}. 

This raises a fundamental question regarding the relevance of local and non-local contributions to the OHE for the orbital signal measured in different types of experiments. For ground-state OM, for example, it is well established that the relationship between this fundamental quantity and the signal measured by spectroscopic techniques such as XMCD or photoemission is not straightforward~\cite{verma2025orbital,stohr1999exploring,van1993spin}. Similarly, establishing a relation between current-induced orbital accumulation at a surface and bulk orbital currents requires careful consideration of surface effects and scattering mechanisms~\cite{go2024local}. When interpreting orbital torque experiments in magnetic heterostructures, it is additionally important to consider the selectivity of the orbital current absorption by the local exchange field of magnetic atoms, which is crucial in mediating specific types of torques~\cite{zeer2025orbital,hayashi2023observation,hayashi2025crystallographic,go2025orbital,fukunaga2023orbital,go2020orbital}.

It was uncovered recently that orbital matching of states at an interface between a magnet and a substrate can be a crucial factor in shaping the torque properties, with certain regions of $k$-space in the electronic structure of the substrate being directly accessible to the ferromagnet while others are not~\cite{go2023long,hayashi2025crystallographic}. In this context, the relevance of various contributions to the OHE discussed here can be strongly $k$-dependent. While in some cases the entire Fermi sea of electrons can contribute fully to the torque, precise control over the crystallinity and structural details of the interface may be employed to suppress one type of OAM current in favor of another. This is exemplified by bcc V, where the OHC extracted from torque measurements agrees with our estimate of the IC contribution, which carries the opposite sign to the ACA prediction~\cite{vijayan2025observation}. This implies that different probing methods and experimental setups may be sensitive to different contributions of OAM. As shown in Fig.~\ref{fig2} for fcc Pt, the sign and magnitude of the LC and IC contributions to the OHC can differ substantially depending on the region of $k$-space. Exploiting these differences for the design and crystalline control of orbital torque presents an exciting experimental challenge for the future.

\section*{Discussion}

\quad In summary, we have introduced a rigorous approach to computing orbital Hall effect which is suitable for implementation from first principles. We showed that the proper treatment of the OAM operator, which reproduces the results of the modern theory of orbital magnetization and is proven to be gauge-covariant,
is key to assessing the role of non-local contributions to the OHC.
Namely, due to its inherent non-locality, the modern Wannier-based OAM operator gives rise to additional terms in orbital quantities such as the OHC. In particular, the so-called ''non-Kubo'' contribution to the OHC, that is entirely absent in all atom-centered treatments and has a purely geometric origin rooted in the gauge structure of the Wannier representation, is shown to be significant in some cases. Our first-principles calculations provide the first systematic values of LC and IC contributions to the OHE in transition metals and a range of other materials. For both the LC and IC parts of the OHC, large contributions arise from the non-local part of the modern OAM operator, and neglecting these non-local contributions 
leads to results very similar to the atom-centered approximation. This underlines the fundamental importance of non-local corrections, while simultaneously revealing the deep connection between the modern Wannier-based framework and the conventional atom-centered viewpoint. The present work raises important questions regarding the interplay between LC and IC, and between local and non-local contributions, in the transport and non-equilibrium dynamics of angular momentum, motivating further theoretical and experimental studies in this area.

\section*{Methods}

\quad All first-principles calculations were performed within the full-potential linearized augmented plane-wave (FLAPW) framework as implemented in v26 of the FLEUR code~\cite{fleurCode,fleurWeb}. Spin-orbit coupling was included self-consistently in all calculations, with the quantization axis oriented along the $z$-direction. Maximally localized Wannier functions were subsequently constructed using the Wannier90 code~\cite{pizzi2020wannier90}, employing standard disentanglement procedures to obtain well-converged Wannier representations for each material.

The OHC was computed using the ORBITRANS code, which implements the Wannier interpolation framework developed in this work and will be published in full detail in a forthcoming contribution~\cite{Orbitrans}. The OHC was evaluated by integrating over a uniform $250 \times 250 \times 250$ $k$-mesh in the full Brillouin zone. A thermal broadening of $k_{\textrm{B}}T = 0.025\,\mathrm{eV}$ was used throughout. To suppress spurious contributions from near-degenerate band crossings at finite temperature, we introduced a small spin Zeeman field of strength $J_{\textrm{Z}} = 10^{-4}\,\mathrm{eV}$, and applied an energy cutoff of $10^{-4}\,\mathrm{eV}$ below which contributions from terms containing the $\overline{J}$ matrix are discarded when the energies of an occupied and an unoccupied band become nearly degenerate. For the computaion of $\overline{J}$ we have used $\eta = 0.025 \,eV$.

The reference atom-centered approximation (ACA) values were computed using the same $k$-mesh and broadening parameters to ensure a consistent comparison. Further details of the derivations and the explicit form of the non-Kubo contribution $\delta\tilde{L}_{\alpha}$ are provided in the Supplementary Material.

\section*{Data availability}
The data that support the findings of this study are available from the corresponding authors upon reasonable request.

\section*{Code availability}
The ORBITRANS code used in this work will be published in a forthcoming contribution. Additional scripts and workflows are available from the corresponding authors upon reasonable request.

\section*{Acknowledgements}

We thank Ivo Souza, Garima Ahuja, Felix L\"upke, Lukasz Plucinski, Aur\'elien Manchon, Daegeun Jo, Peter Oppeneer, and Suik Cheon for fruitful discussions. H.-W. Lee, I. Baek, and H. Lee were financially supported by the National Research Foundation of Korea (NRF) grant funded by the Korean government (MSIT) (No. \textcolor{red}{RS-2024-00356270, RS-2024-00410027}). This work was supported by the J\"ulich Supercomputing Centre (jiff40) and the National Supercomputing Center (KSC-2025-CRE-0068). We also acknowledge support by the Deutsche Forschungsgemeinschaft (DFG) in the framework of TRR 288/2 $-$ 422213477 (Project B06), TRR 173/3 $-$ 268565370 (project A11) and by the EIC Pathfinder OPEN grant 101129641 ``OBELIX''.

\section*{Author contributions}
M.S., I.B., H.L., and D.G. developed the theoretical framework. M.S. performed the calculations. H.-W.L. and Y.M. contributed to the analysis and interpretation of the results. All authors discussed the results and contributed to writing the manuscript.

\section*{Competing interests}
The authors declare no competing interests.
\bibliographystyle{naturemag}
\bibliography{reference}
\end{document}


\maketitle

\footnotetext[1]{These authors contributed equally to this work.}
\footnotetext[2]{\hyperlink{mailto:m.sastges@fz-juelich.de}{m.sastges@fz-juelich.de}}
\footnotetext[3]{\hyperlink{mailto:y.mokrousov@fz-juelich.de}{y.mokrousov@fz-juelich.de}}
\footnotetext[4]{\hyperlink{mailto:dongwookgo@korea.ac.kr}{dongwookgo@korea.ac.kr}}

\renewcommand{\thefootnote}{\arabic{footnote}}

\section*{A. Orbital Angular Momentum Operator}

For deriving the proposed orbital angular momentum (OAM) operator, we adopt a methodology analogous to that employed by Thonhauser \textit{et al.}~\cite{thonhauser2005orbital}. Unlike that work, however, we begin from the following real-space matrix element between Wannier states $\vert n\textbf{R} \rangle$ centered at lattice site $\textbf{R}$:

\begin{equation} \label{AngMomWannierFunc}
\langle n\textbf{R} \vert \hat{\textbf{L}} \vert m\textbf{R} \rangle = \langle n\textbf{R} \vert \hat{\textbf{r}} \times \hat{\textbf{p}} \vert m\textbf{R} \rangle = \langle n\textbf{R} \vert \left( \hat{\textbf{r}} - \textbf{R} \right) \times \hat{\textbf{p}} \vert m\textbf{R} \rangle + \textbf{R} \times \langle n\textbf{R} \vert \hat{\textbf{p}} \vert m\textbf{R} \rangle ,
\end{equation}
which separates the OAM naturally into a term local to the Wannier center and a term proportional to the lattice vector $\textbf{R}$. Following Thonhauser \textit{et al.}~\cite{thonhauser2005orbital}, we divide the crystal into a bulk region and a surface region, positioning the boundary sufficiently deep that the surface region can still be described using bulk Wannier functions~\cite{thonhauser2005orbital,thonhauser2011theory,ceresoli2006orbital}. We assume a total of $N_{\textrm{C}}$ sites, of which $N_{\textrm{B}}$ belong to the bulk and $N_{\textrm{S}}$ to the surface, with the reciprocal space sampled on a $k$-mesh of $N$ points.

In the thermodynamic limit, $N_{\textrm{S}}/N_{\textrm{B}} \rightarrow 0$ and $N_{\textrm{B}} \rightarrow N_{\textrm{C}}$ holds. Under this limit, summing over all sites $\textbf{R}$ shows that the local term $\langle n\textbf{R} \vert (\hat{\textbf{r}} - \textbf{R}) \times \hat{\textbf{p}} \vert m\textbf{R} \rangle$ receives contributions exclusively from the bulk, while the itinerant term $\textbf{R} \times \langle n\textbf{R} \vert \hat{\textbf{p}} \vert m\textbf{R} \rangle$ receives contributions exclusively from the surface~\cite{thonhauser2005orbital,thonhauser2011theory,ceresoli2006orbital}. This is the real-space origin of the local-circulation (LC) and itinerant-circulation (IC) decomposition of the OAM. The surface contribution to the local term becomes negligible because it takes the same value at every site while bulk sites vastly outnumber surface sites. The bulk contribution to the itinerant term vanishes due to the following identity when summing over all bulk sites:

\begin{equation} \label{BulkZero}
\begin{split}
\frac{1}{N_{\textrm{C}}}\sum_{\textbf{R}} \textbf{R} \times \langle n\textbf{R} \vert \hat{\textbf{p}} \vert m\textbf{R} \rangle
= & \frac{1}{NN_{\textrm{C}}} \sum_{\textbf{R}\textbf{k}\textbf{k}'} \left[ \textbf{R} e^{i(\textbf{k}-\textbf{k}') \cdot \textbf{R}} \right] \times \langle u_{n\textbf{k}} \vert \hat{\textbf{p}} \vert u_{m\textbf{k}'} \rangle e^{i(\textbf{k}'-\textbf{k}) \cdot \textbf{r}} \\
= & \frac{1}{NN_{\textrm{C}}}\sum_{\textbf{R}\textbf{k}\textbf{k}'} \left[ -i \partial_{\textbf{k}} e^{i(\textbf{k}-\textbf{k}') \cdot \textbf{R}} \right] \times \langle u_{n\textbf{k}}\vert \hat{\textbf{p}} \vert u_{m\textbf{k}'} \rangle e^{i(\textbf{k}'-\textbf{k}) \cdot \textbf{r}} \\
= & -\frac{i}{N}\sum_{\textbf{k}\textbf{k}'} \partial_{\textbf{k}} \left[ \frac{1}{N_{\textrm{C}}}\sum_{\textbf{R}} e^{i(\textbf{k}-\textbf{k}') \cdot \textbf{R}} \right] \times \langle u_{n\textbf{k}} \vert \hat{\textbf{p}} \vert u_{m\textbf{k}'} \rangle e^{i(\textbf{k}'-\textbf{k}) \cdot \textbf{r}} \\
= & -i\sum_{\textbf{k}\textbf{k}'} \partial_{\textbf{k}} \delta_{\textbf{k}\textbf{k}'} \times \langle u_{n\textbf{k}} \vert \hat{\textbf{p}} \vert u_{m\textbf{k}'} \rangle e^{i(\textbf{k}'-\textbf{k}) \cdot \textbf{r}} = 0 .
\end{split}
\end{equation}
The contribution of the LC term from the bulk region can be brought into reciprocal space. Using $(\hat{\textbf{r}}-\textbf{R}) \times (\hat{\textbf{r}}-\textbf{R}) = 0$, which holds throughout the bulk~\cite{thonhauser2005orbital,thonhauser2011theory,ceresoli2006orbital}, one obtains:

\begin{equation} \label{LCThon}
\begin{split}
\frac{1}{N_{\textrm{C}}} \sum_{\textbf{R}} \langle n\textbf{R} \vert \left( \hat{\textbf{r}} - \textbf{R} \right) \times \hat{\textbf{p}} \vert m\textbf{R} \rangle = & \frac{ie}{2\mu_{\textrm{B}}N_{\textrm{C}}} \sum_{\textbf{R}} \langle n\textbf{R} \vert \left(\hat{\textbf{r}} - \textbf{R}\right) \times \left[ \hat{\mathcal{H}}, \left(\hat{\textbf{r}} - \textbf{R}\right) \right] \vert m\textbf{R} \rangle \\ = & \frac{ie}{2\mu_{\textrm{B}}N_{\textrm{C}}} \sum_{\textbf{R}} \langle n\textbf{R} \vert \left(\hat{\textbf{r}} - \textbf{R}\right) \times \hat{\mathcal{H}} \left(\hat{\textbf{r}} - \textbf{R}\right) \vert m\textbf{R} \rangle \\ = & \frac{ie}{2\mu_{\textrm{B}}NN_{\textrm{C}}} \sum_{\textbf{Rkk}'} \langle u_{n\textbf{k}} \vert e^{-i\textbf{k}\cdot\left(\hat{\textbf{r}} - \textbf{R} \right)} \left(\hat{\textbf{r}} - \textbf{R}\right) \times \hat{\mathcal{H}} \left(\hat{\textbf{r}} - \textbf{R}\right) e^{i\textbf{k}'\cdot\left(\hat{\textbf{r}} - \textbf{R} \right)} \vert u_{m\textbf{k}'} \rangle \\ = & \frac{ie}{2\mu_{\textrm{B}}N} \sum_{\textbf{kk}'} \langle u_{n\textbf{k}} \vert \partial_{\textbf{k}} \left[ e^{-i \textbf{k} \cdot \hat{\textbf{r}}} \times \hat{\mathcal{H}} \partial_{\textbf{k}'} \left[ e^{i\textbf{k}'\cdot\hat{\textbf{r}}} \frac{1}{N_{C}} \sum_{\textbf{R}} e^{i\left(\textbf{k} - \textbf{k}'\right)\cdot \textbf{R}} \right]\right]\vert u_{m\textbf{k}'} \rangle \\ = & \frac{ie}{2\mu_{\textrm{B}}N} \sum_{\textbf{kk}'} \langle u_{n\textbf{k}} \vert \partial_{\textbf{k}} \left[ e^{-i \textbf{k} \cdot \hat{\textbf{r}}} \times \hat{\mathcal{H}} \partial_{\textbf{k}'} \left[ e^{i\textbf{k}' \cdot\hat{\textbf{r}}} \delta_{\textbf{kk}'} \right]\right]\vert u_{m\textbf{k}'} \rangle \\ = & \frac{ie}{2\mu_{\textrm{B}}N} \sum_{\textbf{k}} \langle u_{n\textbf{k}} \vert \hat{\textbf{r}} \times e^{-i\textbf{k} \cdot\hat{\textbf{r}}} \hat{\mathcal{H}} e^{i\textbf{k} \cdot\hat{\textbf{r}}} \hat{\textbf{r}} \vert u_{m\textbf{k}} \rangle \\ = & \frac{ie}{2\mu_{\textrm{B}}N} \sum_{\textbf{k}} \langle \partial_{\textbf{k}} u_{n\textbf{k}} \vert \times \hat{H}_{\textbf{k}} \vert \partial_{\textbf{k}} u_{m\textbf{k}} \rangle .
\end{split}
\end{equation}

To evaluate the surface contribution in reciprocal space, we introduce the macroscopic surface current matrix following Refs.~\cite{thonhauser2005orbital,thonhauser2011theory,ceresoli2006orbital}. Considering a face of a large box whose surface normal points along the positive $x$-axis, and treating the Wannier center positions as continuous, the macroscopic surface current matrix reads:

\begin{equation} \label{MacSurfCurr}
\begin{split}
\textbf{I}_{nm,+x} & = -\frac{e}{S} \sum_{\textbf{R}}^{V_{\textrm{S}}} \langle n\textbf{R} \vert \hat{\textbf{v}} \vert m\textbf{R} \rangle \\& = -\frac{ie}{\hbar S} \sum_{\textbf{R}}^{V_{\textrm{S}}} \langle n\textbf{R} \vert \left[ \hat{\mathcal{H}} \hat{\textbf{r}} - \hat{\textbf{r}} \hat{\mathcal{H}} \right] \vert m\textbf{R} \rangle \\& = -\frac{ie}{\hbar S} \sum_{\textbf{R}}^{V_{\textrm{S}}} \sum_{\textbf{R}'}^{V_{\textrm{B}}} \sum_{n'} \left[ \langle n\textbf{R} \vert \hat{\mathcal{H}} \vert n'\textbf{R}' \rangle \langle n'\textbf{R}' \vert \hat{\textbf{r}} \vert m\textbf{R} \rangle - \langle n\textbf{R} \vert \hat{\textbf{r}} \vert n'\textbf{R}' \rangle \langle n'\textbf{R}' \vert \hat{\mathcal{H}} \vert m\textbf{R} \rangle \right] \\& = -\frac{ie}{\hbar S} \sum_{\textbf{R}}^{V_{\textrm{S}}}\sum_{\textbf{R}'}^{V_{\textrm{B}}}\sum_{n'} \textbf{v}_{\langle n\textbf{R},n'\textbf{R}',m\textbf{R} \rangle} \\& = -\frac{ie}{\hbar S} \sum_{\textbf{R}}^{V_{\textrm{S}}} \sum_{\textbf{R}'}^{V_{\textrm{B}}} \sum_{n'}\textbf{v}_{\langle n\textbf{0},n'\textbf{R}'-\textbf{R},m\textbf{0} \rangle}  \\& = -\frac{ie}{2\hbar V} \sum_{n'\textbf{R}}\left( \textbf{R}\cdot \hat{\textbf{n}}_{x} \right) \textbf{v}_{\langle n\textbf{0},n'\textbf{R},m\textbf{0} \rangle} ,
\end{split}
\end{equation}
where $S$ is the surface area, $V_{\textrm{B}}$ and $V_{\textrm{S}}$ denote the volumes of the bulk and surface regions respectively, $\cdot n_{\textrm{S}}$ is the surface normal, here specifically $\hat{\textbf{n}}_{x}$, and $V$ is the volume of a single crystal unit cell~\cite{ceresoli2006orbital}. The final step of Eq.~(\ref{MacSurfCurr}) follows from a counting argument: for a given value of $\left(\textbf{R}' - \textbf{R}\right) \cdot n_{\textrm{S}} > 0$, the double sum over surface and bulk positions yields exactly $\frac{ \left( \textbf{R}' - \textbf{R}\right) \cdot n_{\textrm{S}} S}{V}$ contributing pairs~\cite{thonhauser2005orbital,thonhauser2011theory,ceresoli2006orbital}. Additionally, the factor of $\frac{1}{2}$ extends the summation over $\textbf{R}$ to the whole volume~\cite{ceresoli2006orbital}. Adding contributions from all faces, the itinerant term evaluates to~\cite{thonhauser2005orbital,thonhauser2011theory,ceresoli2006orbital}:

\begin{equation} \label{Curr3}
\begin{split}
\frac{1}{N_{\textrm{C}}} \sum_{\textbf{R}}^{N_{\textrm{S}}} \textbf{R} \times \langle n\textbf{R} \vert 
\hat{\textbf{p}} \vert m\textbf{R} \rangle  & = -\frac{ie}{4\mu_{\textrm{B}} N_{\textrm{C}}}\sum_{n'\beta\textbf{R}} R_{\beta}  \textbf{v}_{\langle n\textbf{0},n'\textbf{R},m\textbf{0} \rangle} \times \hat{\textbf{n}}_{\beta} \\& = \frac{ie}{4\mu_{\textrm{B}} N_{\textrm{C}}}\sum_{n'\textbf{R}} \textbf{R} \times \textbf{v}_{\langle n\textbf{0},n'\textbf{R},m\textbf{0} \rangle} .
\end{split}
\end{equation}
This expression is fully determined by bulk Wannier functions and is therefore independent of surface details. Bringing it into reciprocal space using the relation

\begin{equation} \label{RelH2}
\begin{split}
\sum_{\textbf{RR}'} \left( \textbf{R} - \textbf{R}' \right) \times \langle n\textbf{R}\vert \hat{\mathcal{H}} \vert n'\textbf{R}' \rangle \langle n'\textbf{R}'\vert \hat{\textbf{r}} \vert m\textbf{R} \rangle  = & \frac{1}{N^{2}}\sum_{\textbf{k}\textbf{k}'\textbf{k}''\textbf{k}'''\textbf{RR}'} \langle u_{n\textbf{k}} \vert e^{-i\textbf{k}\cdot \hat{\textbf{r}}} \hat{\mathcal{H}} e^{i\textbf{k}' \cdot \hat{\textbf{r}} }\vert u_{n'\textbf{k}'} \rangle \\& \langle u_{n'\textbf{k}''} \vert e^{-i\textbf{k}''\cdot \hat{\textbf{r}}} \left( \hat{\textbf{r}}-\textbf{R}'\right) \times \left( \hat{\textbf{r}}-\textbf{R}\right) e^{i\textbf{k}'''\cdot \hat{\textbf{r}}} \vert u_{m\textbf{k}'''} \rangle \\& e^{i\left(\textbf{k} - \textbf{k}''' \right)\cdot\textbf{R}} e^{-i\left(\textbf{k}' - \textbf{k}'' \right)\cdot \textbf{R}'} \\ = & \frac{N_{\textrm{C}}^{2}}{N}\sum_{\textbf{k}}\langle u_{n\textbf{k}} \vert \hat{H}_{\textbf{k}} \vert u_{n'\textbf{k}} \rangle \langle \partial_{\textbf{k}}u_{n'\textbf{k}} \vert \times \vert \partial_{\textbf{k}}u_{m\textbf{k}} \rangle \\ = & \frac{N_{\textrm{C}}^{2}}{N}\sum_{\textbf{k}} \mathbb{H}_{nn',\textbf{k}} \langle \partial_{\textbf{k}}u_{n'\textbf{k}} \vert \times \vert \partial_{\textbf{k}}u_{m\textbf{k}} \rangle ,
\end{split}
\end{equation}
yields the final expression for the itinerant term:

\begin{equation} \label{Curr4}
\begin{split}
\frac{1}{N_{\textrm{C}}} \sum_{\textbf{R}}^{N_{\textrm{S}}} \textbf{R} \times \langle n\textbf{R} \vert 
\hat{\textbf{p}} \vert m\textbf{R} \rangle = & \frac{ie}{4\mu_{\textrm{B}} N_{\textrm{C}}^{2}}\sum_{n'\textbf{R}\textbf{R'}} \left(\textbf{R} - \textbf{R}' \right) \times \textbf{v}_{\langle n\textbf{R}',n'\textbf{R},m\textbf{R}' \rangle} \\ = & \frac{ie}{4\mu_{\textrm{B}}N} \sum_{n'\textbf{k}} \Big[ \mathbb{H}_{nn',\textbf{k}} \langle \partial_{\textbf{k}}u_{n'\textbf{k}} \vert \times \vert \partial_{\textbf{k}}u_{m\textbf{k}} \rangle + \langle \partial_{\textbf{k}}u_{n\textbf{k}} \vert \times \vert \partial_{\textbf{k}}u_{n'\textbf{k}} \rangle \mathbb{H}_{n'm,\textbf{k}} \Big] .
\end{split}
\end{equation}

Combining the local and itinerant terms, and specializing to the case where the Bloch states are eigenstates of the Hamiltonian, yields the $k$-resolved OAM operator matrix element:

\begin{equation} \label{AngularMomMethod2}
\langle u_{n\textbf{k}}^{\textrm{H}} \vert \hat{\textbf{L}} \vert u_{m\textbf{k}}^{\textrm{H}} \rangle = \frac{ie}{2\mu_{B}} \langle \partial_{\textbf{k}} u_{n\textbf{k}}^{\textrm{H}} \vert \times \left( \hat{H}_{\textbf{k}} + \frac{\varepsilon_{n\textbf{k}} + \varepsilon_{m\textbf{k}} }{2} - 2 \mu \right) \vert \partial_{\textbf{k}}u_{m\textbf{k}}^{\textrm{H}} \rangle ,
\end{equation}

where the chemical potential $\mu$ is included to account for shifts of the energy zero. However, this expression is not gauge-covariant and further modifications need to be applied to describe the OAM in a proper way. 

\section*{B. Covariant Derivative}

As discussed in the main manuscript, expressing Eq.~(\ref{AngularMomMethod2}) in terms of Berry-type matrices introduces the Hermitian matrix $\overline{J}$, which generates uncontrolled poles whenever two bands become degenerate. In the zero-temperature limit, this issue is resolved by introducing a covariant derivative $D_{\textbf{k}}$, which simultaneously renders the OAM operator gauge-covariant and suppresses the numerical instabilities.

We define the covariant derivative to act differently on occupied and inner unoccupied states: derived occupied states are projected onto the entire unoccupied space, while derived inner unoccupied states are projected onto both the occupied and outer spaces. We refer to this as the occupation-weighted covariant derivative:

\begin{equation} \label{CovDerOccUnocc}
\begin{split}
\vert D_{\textbf{k}} u_{n\textbf{k}}^{\textrm{H}} \rangle
= & \hat{Q}_{\textbf{k}} \vert \partial_{\textbf{k}}u_{n\textbf{k}}^{\textrm{H}} \rangle f_{n}^{\textrm{H}} + \left( \hat{P}_{\textbf{k}} + \hat{\mathbb{Q}}_{\textbf{k}} \right) \vert \partial_{\textbf{k}}u_{n\textbf{k}}^{\textrm{H}} \rangle g_{n}^{\textrm{H}} \\
= & \left(\hat{\mathbb{1}} - \hat{P}_{\textbf{k}}\right) \vert \partial_{\textbf{k}}u_{n\textbf{k}}^{\textrm{H}} \rangle f_{n}^{\textrm{H}} + \left( \hat{\mathbb{1}} - \hat{Q}_{\textrm{in},\textbf{k}} \right) \vert \partial_{\textbf{k}}u_{n\textbf{k}}^{\textrm{H}} \rangle g_{n}^{\textrm{H}} \\
= & \vert \partial_{\textbf{k}} u_{n\textbf{k}}^{\textrm{H}} \rangle + i \sum_{m} \vert u_{m\textbf{k}}^{\textrm{H}} \rangle \left( f_{m}^{\textrm{H}} A_{mn} f_{n}^{\textrm{H}} + g_{m}^{\textrm{H}} A_{mn} g_{n}^{\textrm{H}} \right) ,
\end{split}
\end{equation}
where $\hat{P}_{\textbf{k}}$ and $\hat{Q}_{\textbf{k}} = \hat{\mathbb{1}} - \hat{P}_{\textbf{k}}$ project onto the occupied and total unoccupied spaces, $\hat{Q}_{\textrm{in},\textbf{k}}$ projects onto the inner (Wannier-spanned) unoccupied space, $\hat{\mathbb{Q}}_{\textbf{k}}$ projects onto the outer space not spanned by the Wannier basis, and $f$, $g$ are the corresponding occupation functions satisfying $f + g = 1$ and $fg = gf = 0$. Introducing the covariant derivative in this way eliminates the gauge dependence of the OAM operator matrix elements. Substituting into Eq.~(\ref{AngularMomMethod2}) yields the modern OAM operator:

\begin{equation} \label{AngularMomMethod3}
\langle u_{n\textbf{k}}^{\textrm{H}} \vert \hat{\textbf{L}} \vert u_{m\textbf{k}}^{\textrm{H}} \rangle = \frac{ie}{2\mu_{B}} \langle D_{\textbf{k}} u_{n\textbf{k}}^{\textrm{H}} \vert \times \left( \hat{H}_{\textbf{k}} + \frac{\varepsilon_{n\textbf{k}} + \varepsilon_{m\textbf{k}} }{2} - 2 \varepsilon_{\textrm{F}} \right) \vert D_{\textbf{k}}u_{m\textbf{k}}^{\textrm{H}} \rangle .
\end{equation}
Expanding in terms of Berry-type matrices yields the full expression:

\begin{equation}
\label{CovariantAngularMomentumFull}
\begin{split}
\langle u_{n\textbf{k}}^{\textrm{H}} \vert \hat{\textbf{L}} \vert u_{m\textbf{k}}^{\textrm{H}} \rangle = & \frac{e}{2\mu_{B}} \Big[ U^{\dagger} \big( \mathbb{\Lambda} - i f \mathbb{A} \times f \mathbb{B} - i g \mathbb{A} \times g \mathbb{B} - i \mathbb{B}^{\dagger} \times f \mathbb{A} f - i \mathbb{B}^{\dagger} \times g \mathbb{A} g \\
& + i f \mathbb{A} \times f \mathbb{H} f \mathbb{A} f + i g \mathbb{A} \times g \mathbb{H} g \mathbb{A} g
+ i f J \times g \mathbb{B} + i \mathbb{B}^{\dagger} \times g J f \\
& - i f J \times g \mathbb{H} g \mathbb{A} g - i g \mathbb{A} \times g \mathbb{H} g J f + i g J \times f \mathbb{H} f \mathbb{A} g + i g \mathbb{A} \times f \mathbb{H} f J g \\
& + i f J \times g \mathbb{H} g J f + i g J \times f \mathbb{H} f J g \\
& + \frac{1}{2}\big\{ \mathbb{H} - 2\varepsilon_{F} ,\: \mathbb{\Omega} - i \mathbb{A} \times \mathbb{A} + i f \mathbb{A} \times g \mathbb{A} f + i g \mathbb{A} \times f \mathbb{A} g \\
& + i f J \times g \mathbb{A} f + i g J \times f \mathbb{A} g + i f \mathbb{A} \times g J f + i g \mathbb{A} \times f J g \\
& + i f J \times g J f + i g J \times f J g \big\}\big) U \Big]_{nm} .
\end{split}
\end{equation}

As is evident from this expression, every $J$-matrix appears exclusively sandwiched between occupation functions of different subspaces (i.e.\ between $f$ and $g$ or vice versa), which suppresses the emergence of poles exactly in the zero-temperature limit and serves as a valid approximation at low temperatures. At finite temperatures, residual poles may arise; these are regularized by introducing a broadening $\eta$ in the denominator of $\overline{J}$:

\begin{equation} \label{JBroadened}
\overline{J}_{nm} = i \left[ U^{\dagger} \partial_{\textbf{k}} U \right]_{nm} = \frac{i\left[ U^{\dagger} \left(\partial_{\textbf{k}}\mathbb{H}\right) U \right]_{nm}}{\varepsilon_{m\textbf{k}} - \varepsilon_{n\textbf{k}} + i\eta} .
\end{equation}
Note that this renders $\overline{J}$ non-Hermitian; consequently, every $\overline{J}$ appearing on the left side of a vector product in Eq.~(\ref{CovariantAngularMomentumFull}) must be replaced by its Hermitian conjugate $\overline{J}^{\dagger}$.

Equation~(\ref{CovariantAngularMomentumFull}) can be decomposed into its LC contribution,

\begin{equation} \label{CovariantAngularMomentumFullLC}
\begin{split}
\langle u_{n\textbf{k}}^{\textrm{H}} \vert \hat{\textbf{L}}^{\textrm{LC}} \vert u_{m\textbf{k}}^{\textrm{H}} \rangle = & \frac{e}{2\mu_{B}} \Big[ U^{\dagger} \Big( \mathbb{\Lambda} - i f \mathbb{A} \times f \mathbb{B} - i g \mathbb{A} \times g \mathbb{B} - i \mathbb{B}^{\dagger} \times f \mathbb{A} f - i \mathbb{B}^{\dagger} \times g \mathbb{A} g \\
& + i f \mathbb{A} \times f \mathbb{H} f \mathbb{A} f + i g \mathbb{A} \times g \mathbb{H} g \mathbb{A} g
+ i f J \times g \mathbb{B} + i \mathbb{B}^{\dagger} \times g J f \\
& - i f J \times g \mathbb{H} g \mathbb{A} g - i g \mathbb{A} \times g \mathbb{H} g J f + i g J \times f \mathbb{H} f \mathbb{A} g + i g \mathbb{A} \times f \mathbb{H} f J g \\
& + i f J \times g \mathbb{H} g J f + i g J \times f \mathbb{H} f J g \\
& - \varepsilon_{F} \big( \mathbb{\Omega} - i \mathbb{A} \times \mathbb{A} + i f \mathbb{A} \times g \mathbb{A} f + i g \mathbb{A} \times f \mathbb{A} g \\
& + i f J \times g \mathbb{A} f + i g J \times f \mathbb{A} g + i f \mathbb{A} \times g J f + i g \mathbb{A} \times f J g \\
& + i f J \times g J f + i g J \times f J g \big) \Big) U \Big]_{nm} ,
\end{split}
\end{equation}
and its IC contribution,

\begin{equation} \label{CovariantAngularMomentumFullIC}
\begin{split}
\langle u_{n\textbf{k}}^{\textrm{H}} \vert \hat{\textbf{L}}^{\textrm{IC}} \vert u_{m\textbf{k}}^{\textrm{H}} \rangle = & \frac{e}{4\mu_{B}} \Big[ U^{\dagger} \big\{ \mathbb{H} - \varepsilon_{F} ,\: \mathbb{\Omega} - i \mathbb{A} \times \mathbb{A} + i f \mathbb{A} \times g \mathbb{A} f + i g \mathbb{A} \times f \mathbb{A} g \\
& + i f J \times g \mathbb{A} f + i g J \times f \mathbb{A} g + i f \mathbb{A} \times g J f + i g \mathbb{A} \times f J g \\
& + i f J \times g J f + i g J \times f J g \big\} U \Big]_{nm} .
\end{split}
\end{equation}

A key consistency check is that the modern OAM operator reproduces the ground-state orbital magnetization of Lopez and collaborators~\cite{lopez2012wannier} exactly. This is demonstrated by multiplying Eq.~(\ref{CovariantAngularMomentumFull}) by the Fermi occupation function $f$ and tracing:

\begin{equation} \label{MagFull}
\begin{split}
\textbf{M}\left( \textbf{k} \right) \sim & \mathrm{tr}\left(f\hat{\textbf{L}}\right) \\ \sim & \mathrm{tr}\Big[ f\mathbb{\Lambda}f - i f \mathbb{A} \times f \mathbb{B}f - i f \mathbb{B}^{\dagger} \times f \mathbb{A} f + i f \mathbb{A} \times f \mathbb{H} f \mathbb{A} f + i f J \times g \mathbb{B} f + i f \mathbb{B}^{\dagger} \times g J f \\& + i f J \times g \mathbb{H} g J f + \big( f \mathbb{H} - 2\varepsilon_{F} \big) \big( f\mathbb{\Omega}f - i f\mathbb{A} \times \mathbb{A}f + i f \mathbb{A} \times g \mathbb{A} f + i f J \times g \mathbb{A} f + if \mathbb{A} \times g J f \big) \Big] \\ = & \mathrm{tr}\Big[ f\mathbb{\Lambda}f + i f J \times g \mathbb{B} f + i f \mathbb{B}^{\dagger} \times g J f + i f J \times g \mathbb{H} g J f + \big( f\mathbb{H} - 2\varepsilon_{F} \big) \big( f\mathbb{\Omega}f + i f J \times g \mathbb{A} f + if \mathbb{A} \times g J f \big) \Big].
\end{split}
\end{equation}
with the LC and IC parts given separately by

\begin{equation} \label{MagLC}
\begin{split}
\textbf{M}^{\textrm{LC}}\left( \textbf{k} \right) \sim & \mathrm{tr}\left(f\hat{\textbf{L}}^{\textrm{LC}}\right) \\ \sim & \mathrm{tr}\Big[ f\mathbb{\Lambda}f - i f \mathbb{A} \times f \mathbb{B}f - i f \mathbb{B}^{\dagger} \times f \mathbb{A} f + i f \mathbb{A} \times f \mathbb{H} f \mathbb{A} f + i f J \times g \mathbb{B} f + i f \mathbb{B}^{\dagger} \times g J f \\& + i f J \times g \mathbb{H} g J f - \varepsilon_{F} \big( f\mathbb{\Omega}f - i f\mathbb{A} \times \mathbb{A}f + i f \mathbb{A} \times g \mathbb{A} f + i f J \times g \mathbb{A} f + if \mathbb{A} \times g J f \big) \Big]\\ = & \mathrm{tr}\Big[ f\mathbb{\Lambda}f + i f\mathbb{H}f\mathbb{A}\times f\mathbb{A}f + i f J \times g \mathbb{B} f + i f \mathbb{B}^{\dagger} \times g J f + i f J \times g \mathbb{H} g J f \\& - \varepsilon_{F} \big( f\mathbb{\Omega}f + i f J \times g \mathbb{A} f + if \mathbb{A} \times g J f \big) \Big] ,
\end{split}
\end{equation}

\begin{equation} \label{MagIC}
\begin{split}
\textbf{M}^{\textrm{IC}}\left( \textbf{k} \right) \sim & \mathrm{tr}\left(f\hat{\textbf{L}}^{\textrm{IC}}\right) \\ \sim & \mathrm{tr}\Big[ \big( f\mathbb{H} - \varepsilon_{F} \big) \big( f\mathbb{\Omega}f - i f\mathbb{A} \times \mathbb{A}f + i f \mathbb{A} \times g \mathbb{A} f + i f J \times g \mathbb{A} f + if \mathbb{A} \times g J f \big) \Big] \\ = & \mathrm{tr}\Big[ \big( f\mathbb{H} - \varepsilon_{F} \big)\big( f\mathbb{\Omega}f + i f J \times g \mathbb{A} f + if \mathbb{A} \times g J f \big) - i f\mathbb{H}f\mathbb{A}\times f\mathbb{A}f \Big] ,
\end{split}
\end{equation}
These reproduce exactly the LC and IC orbital magnetization expressions of Lopez \textit{et al.}~\cite{lopez2012wannier}. To arrive at these results, the following identities are used~\cite{lopez2012wannier}:

\begin{equation}
1 = f + g, \quad fg = gf = 0, \quad f\mathbb{H} = \mathbb{H}f = f\mathbb{H}f, \quad f\mathbb{B} = f\mathbb{H}f\mathbb{A}, \quad \mathrm{tr}\big[ f \mathbb{A} \times f \mathbb{A} f \big] = 0 .
\end{equation}
Note that moving a matrix across a vector product introduces a sign change, as exploited in the last identity.

We further verify gauge covariance of the occupation-weighted covariant derivative under a $U(1)$ gauge transformation, parameterized by a $k$-dependent phase $\Phi$:

\begin{equation} \label{CovarianceOfGCDerivative}
\begin{split}
\vert D_{\textbf{k}} u_{n\textbf{k}} \rangle = & \vert \partial_{\textbf{k}} u_{n\textbf{k}} \rangle + i \sum_{m} \vert u_{m\textbf{k}} \rangle \left( f_{m} A_{mn} f_{n} + g_{m} A_{mn} g_{n} \right) \\ = & \left[ \vert \partial_{\textbf{k}} u'_{n\textbf{k}} \rangle + i \vert u'_{n\textbf{k}} \rangle \left(\partial_{\textbf{k}}\Phi\right) + i \sum_{m} \vert u'_{m\textbf{k}} \rangle \left( f_{m} A_{mn} f_{n} + g_{m} A_{mn} g_{n} \right) \right] e^{i\Phi} \\ = & \left[ \vert \partial_{\textbf{k}} u'_{n\textbf{k}} \rangle + i \sum_{m} \vert u'_{m\textbf{k}} \rangle \left( f_{m} \left( A_{mn} + \partial_{\textbf{k}}\Phi \mathbb{1}_{mn}\right) f_{n} + g_{m} \left( A_{mn} + \partial_{\textbf{k}}\Phi \mathbb{1}_{mn} \right) g_{n} \right) \right] e^{i\Phi} \\ = & \left[ \vert \partial_{\textbf{k}} u'_{n\textbf{k}} \rangle + i \sum_{m} \vert u'_{m\textbf{k}} \rangle \left( f_{m} A'_{mn} f_{n} + g_{m} A'_{mn} g_{n} \right) \right] e^{i\Phi} \\ = & \vert D_{\textbf{k}} u'_{n\textbf{k}} \rangle e^{i\Phi} .
\end{split}
\end{equation}
using $1 = f + g$ and the standard Berry connection transformation $A' = A + \partial_{\textbf{k}}\Phi$.

\section*{C. Projector Representation}

\begin{figure}[htbp]
\centering
\includegraphics[angle=0, width=0.35\textwidth]{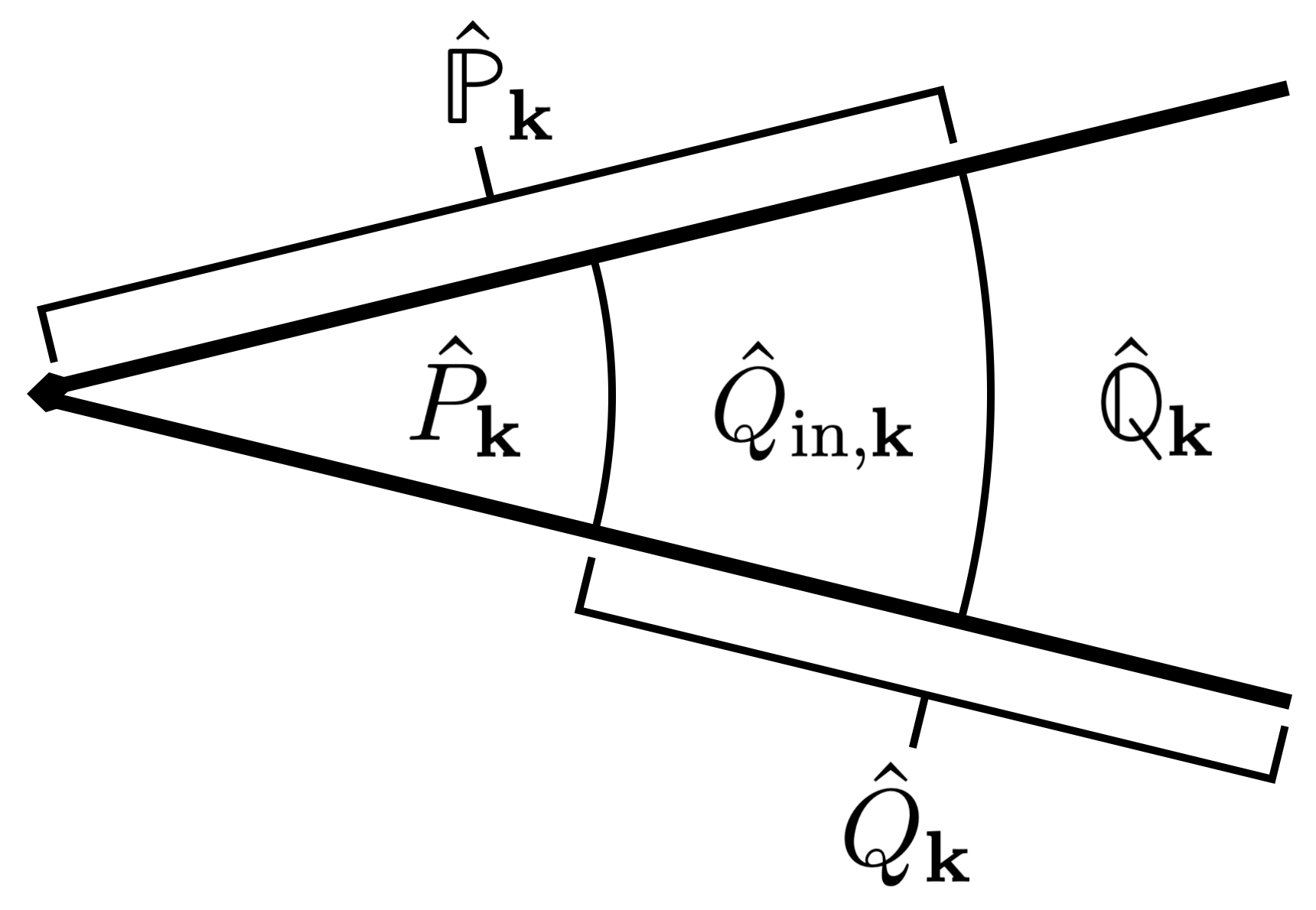}
\caption{Decomposition of the full Hilbert space into the subspaces spanned by the projectors $\hat{P}_{\textbf{k}}$ (occupied), $\hat{Q}_{\textrm{in},\textbf{k}}$ (inner unoccupied), and $\hat{\mathbb{Q}}_{\textbf{k}}$ (outer), following Ref.~\cite{lopez2012wannier}.}
\label{Space}
\end{figure}

The complete gauge covariance of the modern OAM operator is demonstrated most directly by expressing it in terms of projectors onto the occupied ($\hat{P}_{\textbf{k}}$), inner unoccupied ($\hat{Q}_{\textrm{in},\textbf{k}}$), and outer ($\hat{\mathbb{Q}}_{\textbf{k}}$) subspaces (see Fig.~\ref{Space}). Since projectors are manifestly gauge-independent and the Hamiltonian transforms covariantly, any expression built from these objects alone is automatically gauge-covariant. The modern OAM operator admits the following projector form, which closely parallels the ground-state OM expressions of Lopez and co-workers~\cite{lopez2012wannier}:

\begin{equation} \label{ProjOAM}
\begin{split}
\hat{\textbf{L}} = & \frac{ie}{2\mu_{\textrm{B}}} \bigg[ \left( \partial_{\textbf{k}} \hat{P}_{\textbf{k}} \right) \hat{Q}_{\textbf{k}} \times \hat{H}_{\textbf{k}} \hat{Q}_{\textbf{k}} \left( \partial_{\textbf{k}} \hat{P}_{\textbf{k}} \right) + \left( \partial_{\textbf{k}} \hat{Q}_{\textrm{in},\textbf{k}} \right) \left( \hat{\mathbb{1}} - \hat{Q}_{\textrm{in},\textbf{k}} \right) \times \hat{H}_{\textbf{k}} \left( \hat{\mathbb{1}} - \hat{Q}_{\textrm{in},\textbf{k}} \right) \left( \partial_{\textbf{k}} \hat{Q}_{\textrm{in},\textbf{k}} \right) \\
& + \left( \partial_{\textbf{k}} \hat{P}_{\textbf{k}} \right) \hat{Q}_{\textbf{k}} \times \hat{H}_{\textbf{k}} \left( \hat{\mathbb{1}} - \hat{Q}_{\textrm{in},\textbf{k}} \right) \left( \partial_{\textbf{k}} \hat{Q}_{\textrm{in},\textbf{k}} \right) + \left( \partial_{\textbf{k}} \hat{Q}_{\textrm{in},\textbf{k}} \right) \left( \hat{\mathbb{1}} - \hat{Q}_{\textrm{in},\textbf{k}} \right) \times \hat{H}_{\textbf{k}} \hat{Q}_{\textbf{k}} \left( \partial_{\textbf{k}} \hat{P}_{\textbf{k}} \right) \\
& + \frac{1}{2} \Big\{ \hat{H}_{\textbf{k}} - 2\varepsilon_{\textrm{F}},\: \left( \partial_{\textbf{k}} \hat{P}_{\textbf{k}} \right) \times \hat{Q}_{\textbf{k}} \left( \partial_{\textbf{k}} \hat{P}_{\textbf{k}} \right) + \left( \partial_{\textbf{k}} \hat{Q}_{\textrm{in},\textbf{k}} \right) \times \left( \hat{\mathbb{1}} - \hat{Q}_{\textrm{in},\textbf{k}} \right) \left( \partial_{\textbf{k}} \hat{Q}_{\textrm{in},\textbf{k}} \right) \\
& + \left( \partial_{\textbf{k}} \hat{P}_{\textbf{k}} \right) \hat{Q}_{\textbf{k}} \times\left( \hat{\mathbb{1}} - \hat{Q}_{\textrm{in},\textbf{k}} \right) \left( \partial_{\textbf{k}} \hat{Q}_{\textrm{in},\textbf{k}} \right) + \left( \partial_{\textbf{k}} \hat{Q}_{\textrm{in},\textbf{k}} \right) \left( \hat{\mathbb{1}} - \hat{Q}_{\textrm{in},\textbf{k}} \right) \times \hat{Q}_{\textbf{k}} \left( \partial_{\textbf{k}} \hat{P}_{\textbf{k}} \right) \Big\} \bigg] .
\end{split}
\end{equation}
The equivalence of Eq.~(\ref{ProjOAM}) with Eq.~(\ref{CovariantAngularMomentumFull}) follows from the identities~\cite{lopez2012wannier}:
\begin{gather}
\hat{Q}_{\textbf{k}} = \hat{Q}_{\textrm{in},\textbf{k}} + \hat{\mathbb{Q}}_{\textbf{k}} , \quad
\hat{\mathbb{1}} - \hat{Q}_{\textrm{in},\textbf{k}} = \hat{P}_{\textbf{k}} + \hat{\mathbb{Q}}_{\textbf{k}} , \quad
\hat{Q}_{\textbf{k}} \left( \hat{\mathbb{1}} - \hat{Q}_{\textrm{in},\textbf{k}} \right) = \hat{\mathbb{Q}}_{\textbf{k}} , \\
\left( \partial_{\textbf{k}} \hat{P}_{\textbf{k}} \right) \hat{Q}_{\textrm{in},\textbf{k}} = \sum_{nm} \vert u_{n\textbf{k}} \rangle \left[ f \left( i\mathbb{A} \right) g + \left(\partial_{\textbf{k}}f\right) g \right]_{nm} \langle u_{m\textbf{k}} \vert , \quad
\left( \partial_{\textbf{k}} \hat{P}_{\textbf{k}} \right) \hat{\mathbb{Q}}_{\textbf{k}} = \sum_{nm} \vert u_{n\textbf{k}} \rangle f_{nm} \langle \partial_{\textbf{k}} u_{m\textbf{k}} \vert \hat{\mathbb{Q}}_{\textbf{k}} , \\
\left( \partial_{\textbf{k}} \hat{Q}_{\textrm{in},\textbf{k}} \right) \hat{P}_{\textbf{k}} = \sum_{nm} \vert u_{n\textbf{k}} \rangle \left[ g \left( i\mathbb{A} \right) f + \left(\partial_{\textbf{k}}g\right) f \right]_{nm} \langle u_{m\textbf{k}} \vert , \quad
\left( \partial_{\textbf{k}} \hat{Q}_{\textrm{in},\textbf{k}} \right) \hat{\mathbb{Q}}_{\textbf{k}} = \sum_{nm} \vert u_{n\textbf{k}} \rangle g_{nm} \langle \partial_{\textbf{k}} u_{m\textbf{k}} \vert \hat{\mathbb{Q}}_{\textbf{k}} , \\
\partial_{\textbf{k}} f = i \left[ f, J \right] , \quad \partial_{\textbf{k}} g = i \left[ g, J \right] .
\end{gather}

Tracing Eq.~(\ref{ProjOAM}) over all occupied states reproduces the modern ground-state OM expressions of Ref.~\cite{lopez2012wannier}. An equivalent, more transparent projector form that makes the interaction between different subspaces explicit is:

\begin{equation} \label{ProjOAM2}
\begin{split}
\hat{\textbf{L}} = & \frac{ie}{2\mu_{\textrm{B}}} \bigg[ \left( \partial_{\textbf{k}} \hat{\mathbb{P}}_{\textbf{k}} \right) \hat{\mathbb{Q}}_{\textbf{k}} \times \hat{H}_{\textbf{k}} \hat{\mathbb{Q}}_{\textbf{k}} \left( \partial_{\textbf{k}} \hat{\mathbb{P}}_{\textbf{k}} \right) + \left( \partial_{\textbf{k}} \hat{P}_{\textbf{k}} \right) \hat{Q}_{\textrm{in},\textbf{k}} \times \hat{H}_{\textbf{k}} \hat{Q}_{\textrm{in},\textbf{k}} \left( \partial_{\textbf{k}} \hat{P}_{\textbf{k}} \right) + \left( \partial_{\textbf{k}} \hat{Q}_{\textrm{in},\textbf{k}} \right) \hat{P}_{\textbf{k}} \times \hat{H}_{\textbf{k}} \hat{P}_{\textbf{k}} \left( \partial_{\textbf{k}} \hat{Q}_{\textrm{in},\textbf{k}} \right) \\
& + \frac{1}{2} \Big\{ \hat{H}_{\textbf{k}} - 2\varepsilon_{\textrm{F}},\: \left( \partial_{\textbf{k}} \hat{\mathbb{P}}_{\textbf{k}} \right) \times \hat{\mathbb{Q}}_{\textbf{k}} \left( \partial_{\textbf{k}} \hat{\mathbb{P}}_{\textbf{k}} \right) + \left( \partial_{\textbf{k}} \hat{P}_{\textbf{k}} \right) \times \hat{Q}_{\textrm{in},\textbf{k}} \left( \partial_{\textbf{k}} \hat{P}_{\textbf{k}} \right) + \left( \partial_{\textbf{k}} \hat{Q}_{\textrm{in},\textbf{k}} \right) \times \hat{P}_{\textbf{k}} \left( \partial_{\textbf{k}} \hat{Q}_{\textrm{in},\textbf{k}} \right) \Big\} \bigg] .
\end{split}
\end{equation}

\section*{D. Orbital Hall Conductivity}

The orbital Hall conductivity (OHC) is given by

\begin{equation} \label{OHCfromj}
\sigma_{\alpha\beta}^{L_{\gamma}} = \int_{\textrm{BZ}} \frac{d^{d}\textbf{k}}{\left( 2\pi \right)^{d}} \frac{ \partial \langle j_{\alpha}^{L_{\gamma}} \rangle^{\textrm{D}}}{\partial E_{\beta}} \Bigg|_{\textbf{E} \rightarrow \textbf{0}} = \int_{\textrm{BZ}} \frac{d^{d}\textbf{k}}{\left( 2\pi \right)^{d}} \frac{\partial \langle\delta j_{\alpha}^{L_{\gamma}} \rangle^{\textrm{D}}}{\partial E_{\beta}} ,
\end{equation}
obtained by deriving the first-order change in orbital current by the applied electric field~\cite{salemi2022first,busch2023orbital}. The perturbation due to the electric field can be expressed as $\hat{H}_{\textbf{k}}^{(1)} = e \textbf{E}\left(t\right) \cdot \hat{\textbf{r}}$. 

We can evaluate the orbital current by introducing the dynamic gauge

\begin{equation}
\vert u_{n\textbf{k}}^{\textrm{D}} \left( t \right) \rangle = \sum_{m} \vert u_{m\textbf{k}}^{\textrm{H}} \rangle V_{mn}\left(t\right) ,
\end{equation}
with the time-evolution matrix~\cite{haldar2024dynamical,langhoff1972aspects}

\begin{equation} \label{DynamicGaugeMatrix}
V \left( t \right) = \mathcal{T}\textrm{exp} \left( -\frac{i}{\hbar} \int_{t_{0}}^{t} dt'\, e^{\frac{i}{\hbar} \hat{H}_{\textbf{k}}^{(0)} t'} \hat{H}_{\textbf{k}}^{(1)}\left( t' \right) e^{-\frac{i}{\hbar} \hat{H}_{\textbf{k}}^{(0)} t'} \right) ,
\end{equation}
which captures the effect of the perturbation, with $\hat{H}_{\textbf{k}}^{(0)}$ being the unperturbed Bloch Hamiltonian. For our evaluation of the OHC, we define $\textbf{E}\left(t\right) = \textbf{E} e^{\frac{\eta}{\hbar} t}$, with $\eta \rightarrow 0^{+}$ and integrate from $-\infty$ to $0$. We also only consider the first-order of the Dyson series, where the time-evolution matrix can be approximated as $V = \mathbb{1} + \delta V$, which can be evaluated as

\begin{equation} \label{EPertubation}
\begin{split}
\delta V_{nm} = & -\frac{i}{\hbar} e\langle u_{n\textbf{k}}^{\textrm{H}}\vert \textbf{E} \cdot \hat{\textbf{r}} \vert u_{m\textbf{k}}^{\textrm{H}} \rangle \int_{-\infty}^{0} dt'\, e^{\frac{i}{\hbar} \left( \varepsilon_{n} - \varepsilon_{m} -i \eta \right) t'} \\ = & -e\frac{\langle u_{n\textbf{k}}^{\textrm{H}}\vert \textbf{E} \cdot \hat{\textbf{r}} \vert u_{m\textbf{k}}^{\textrm{H}} \rangle}{\varepsilon_{n\textbf{k}} - \varepsilon_{m\textbf{k}} - i\eta}
\\ = & ie\hbar\frac{\langle u_{n\textbf{k}}^{\textrm{H}}\vert \textbf{E} \cdot \hat{\textbf{v}}\vert u_{m\textbf{k}}^{\textrm{H}} \rangle}{\left( \varepsilon_{n\textbf{k}} - \varepsilon_{m\textbf{k}} - i\eta\right) \left( \varepsilon_{n\textbf{k}} - \varepsilon_{m\textbf{k}}\right)} \\ = & \, ie\hbar\frac{\langle u_{n\textbf{k}}^{\textrm{H}}\vert \textbf{E} \cdot \hat{\textbf{v}}\vert u_{m\textbf{k}}^{\textrm{H}} \rangle}{\left( \varepsilon_{n\textbf{k}} - \varepsilon_{m\textbf{k}}\right)^{2} + \eta^{2}} ,
\end{split}
\end{equation}
with $\delta V_{nm} = 0$ for $n = m$. In the last expression of Eq. (\ref{EPertubation}), we expand the expression by $\left(\varepsilon_{n\textbf{k}} - \varepsilon_{m\textbf{k}} + i\eta\right)$ and get rid of the negligible contribution of $i\eta$ in the numerator to arrive at an expression which regularizes the poles that can arise when two energy bands meet at the Fermi energy. Furthermore, this expression contains the Hamiltonian-gauge velocity operator~\cite{go2024first}

\begin{equation}
\hat{\textbf{v}} = \frac{1}{\hbar} \sum_{mm',\textbf{k}} \vert u^{\textrm{H}}_{m\textbf{k}}\rangle \Big[ \delta_{mm'} \partial_{\textbf{k}} \varepsilon_{m\textbf{k}} + i \left( \varepsilon_{m\textbf{k}} - \varepsilon_{m'\textbf{k}} \right) A_{mm'} \Big] \langle u^{\textrm{H}}_{m'\textbf{k}} \vert .
\end{equation}

The non-locality of the modern OAM operator becomes apparent when transforming $\overline{J}$ between the Wannier and dynamic gauges:

\begin{equation} \label{Jchange}
\overline{J}_{\alpha}' = i V^{\dagger} U^{\dagger} \partial_{\alpha} \left(U V\right)
= \overline{J}_{\alpha} + \delta V^{\dagger} \overline{J}_{\alpha} + \overline{J}_{\alpha} \delta V + i \partial_{\alpha} \delta V
= \overline{J}_{\alpha} + \delta V^{\dagger} \overline{J}_{\alpha} + \overline{J}_{\alpha} \delta V + \overline{\tilde{J}}_{\alpha} .
\end{equation}

The appearance of the extra term $\overline{\tilde{J}}_{\alpha} = i\partial_{\alpha}\delta V$ shows that, unlike the velocity operator which transforms locally ($\delta v^{\textrm{D}}_{\alpha} = \delta V^{\dagger} v^{\textrm{H}}_{\alpha} + v^{\textrm{H}}_{\alpha} \delta V$), the non-local OAM operator acquires an additional contribution when brought into the dynamic gauge:

\begin{equation} \label{dLD}
\delta L^{\textrm{D}}_{\alpha} = \delta V^{\dagger} L_{\alpha}^{\textrm{H}} + L_{\alpha}^{\textrm{H}} \delta V + \delta \tilde{L}_{\alpha} .
\end{equation}
Substituting this into the orbital current expression and using the following first-order relation 

\begin{equation}
\begin{split}
& V^{\dagger} V = 1 \\ \Leftrightarrow & 1 + \delta V^{\dagger} + \delta V = 1 \\ \Leftrightarrow & \delta V^{\dagger} + \delta V = 0 ,
\end{split} 
\end{equation}
the total change in orbital current can be written as:

\begin{equation} \label{CurrentKubo}
\begin{split}
\langle \delta j_{\alpha}^{L_{\beta}} \rangle^{\textrm{D}} = & \frac{1}{2} \sum_{nm} f_{n}^{\textrm{H}} \left[ \delta v_{\alpha,nm}^{\textrm{D}} L_{\beta,mn}^{\textrm{H}} + L_{\beta,nm}^{\textrm{H}} \delta v_{\alpha,mn}^{\textrm{D}} + v_{\alpha,nm}^{\textrm{H}} \delta L_{\beta,mn}^{\textrm{D}} + \delta L_{\beta,nm}^{\textrm{D}} v_{\alpha,mn}^{\textrm{H}} \right] \\ = & \frac{1}{2} \sum_{nmn'} f_{n}^{\textrm{H}} \Big[ \delta V_{nn'}^{\dagger} v_{\alpha,n'm}^{\textrm{H}} L_{\beta,mn}^{\textrm{H}} + v_{\alpha,nn'}^{\textrm{H}} \delta V_{n'm} L_{\beta,mn}^{\textrm{H}} + L_{\beta,nm}^{\textrm{H}} \delta V_{mn'}^{\dagger} v_{\alpha,n'n}^{\textrm{H}} + L_{\beta,nm}^{\textrm{H}} v_{\alpha,mn'}^{\textrm{H}} \delta V_{n'n} \\& + v_{\alpha,nm}^{\textrm{H}} \delta V_{mn'}^{\dagger} L_{\beta,n'n}^{\textrm{H}} + v_{\alpha,nm}^{\textrm{H}} L_{\beta,mn'}^{\textrm{H}} \delta V_{n'n} + \delta V_{nn'}^{\dagger} L_{\beta,n'm}^{\textrm{H}} v_{\alpha,mn}^{\textrm{H}} + L_{\beta,nn'}^{\textrm{H}} \delta V_{n'm} v_{\alpha,mn}^{\textrm{H}} \Big] \\& + \frac{1}{2} \sum_{nm} f_{n}^{\textrm{H}} \Big[ v_{\alpha,nm}^{\textrm{H}} \delta \tilde{L}_{\beta,mn} + \delta \tilde{L}_{\beta,nm} v_{\alpha,mn}^{\textrm{H}} \Big] \\ = &  \frac{1}{2} \sum_{nmn'} f_{n}^{\textrm{H}} \Big[ \delta V_{nn'}^{\dagger} v_{\alpha,n'm}^{\textrm{H}} L_{\beta,mn}^{\textrm{H}} + L_{\beta,nm}^{\textrm{H}} v_{\alpha,mn'}^{\textrm{H}} \delta V_{n'n} + v_{\alpha,nm}^{\textrm{H}} L_{\beta,mn'}^{\textrm{H}} \delta V_{n'n} + \delta V_{nn'}^{\dagger} L_{\beta,n'm}^{\textrm{H}} v_{\alpha,mn}^{\textrm{H}} \Big] \\& + \frac{1}{2} \sum_{nm} f_{n}^{\textrm{H}} \Big[ v_{\alpha,nm}^{\textrm{H}} \delta \tilde{L}_{\beta,mn} + \delta \tilde{L}_{\beta,nm} v_{\alpha,mn}^{\textrm{H}} \Big] \\ = &  \sum_{nm} f_{n}^{\textrm{H}} \Big[ \delta V^{\dagger}_{nm} j_{\alpha,mn}^{L_{\beta}} + j_{\alpha,nm}^{L_{\beta}} \delta V_{mn} + \frac{1}{2} \left( v_{\alpha,nm}^{\textrm{H}} \delta \tilde{L}_{\beta,mn} + \delta \tilde{L}_{\beta,nm} v_{\alpha,mn}^{\textrm{H}} \right) \Big] \\ = & \langle \delta j_{\alpha}^{L_{\beta}} \rangle^{\textrm{D}}_{\textrm{Kubo}} + \langle \delta j_{\alpha}^{L_{\beta}} \rangle^{\textrm{D}}_{\textrm{Non-Kubo}} .
\end{split}
\end{equation}
The non-Kubo correction containing $\delta\tilde{\textbf{L}}$ arises entirely from the non-locality of the OAM operator encoded in the $J$-matrices. While in most of the analyzed systems its contribution to the OHC is rather small compared to the Kubo contribution it can still have a significant impact on the result, e.g. in fcc Pt. Written out explicitly, the non-local change in orbital angular momentum $\delta\tilde{\textbf{L}}$ takes the form

\begin{equation}
\label{AdditionalTerm}
\begin{split}
\delta \tilde{\textbf{L}}_{nm} = & \frac{ie}{2\mu_{B}} \Big[ U^{\dagger} \big( f \tilde{J} \times g \mathbb{B} + \mathbb{B}^{\dagger} \times g \tilde{J} f - f \tilde{J} \times g \mathbb{H} g \mathbb{A} g - g \mathbb{A} \times g \mathbb{H} g \tilde{J} f + g \tilde{J} \times f \mathbb{H} f \mathbb{A} g + g \mathbb{A} \times f \mathbb{H} f \tilde{J} g \\
& + f J \times g \mathbb{H} g \tilde{J} f + g J \times f \mathbb{H} f \tilde{J} g + f \tilde{J} \times g \mathbb{H} g J f + g \tilde{J} \times f \mathbb{H} f J g \\
& + \frac{1}{2}\big\{ \mathbb{H} - 2\varepsilon_{F} ,\: f \tilde{J} \times g \mathbb{A} f + g \tilde{J} \times f \mathbb{A} g + f \mathbb{A} \times g \tilde{J} f + g \mathbb{A} \times f \tilde{J} g \\
& + f J \times g \tilde{J} f + g J \times f \tilde{J} g + f \tilde{J} \times g J f + g \tilde{J} \times f J g \big\}\big) U \Big]_{nm} ,
\end{split}
\end{equation}
where the elements of $\overline{\tilde{J}}$ are computed as:

\begin{equation} \label{dLTilde1}
\overline{\tilde{J}}_{\alpha,nm} = -e\hbar\, \partial_{\alpha} \frac{\langle u_{n\textbf{k}}^{\textrm{H}} \vert \textbf{E} \cdot \hat{\textbf{v}} \vert u_{m\textbf{k}}^{\textrm{H}} \rangle}{\left(\varepsilon_{n\textbf{k}} - \varepsilon_{m\textbf{k}}\right)^{2} + \eta^{2}}
= -ie \vert\textbf{E}\vert \frac{\left(\varepsilon_{n\textbf{k}} - \varepsilon_{m\textbf{k}}\right)\partial_{\alpha}A_{\textbf{E},nm} -\partial_{\alpha} \left(\varepsilon_{n\textbf{k}} - \varepsilon_{m\textbf{k}}\right)A_{\textbf{E},nm}}{\left(\varepsilon_{n\textbf{k}} - \varepsilon_{m\textbf{k}}\right)^{2} + \eta^{2}} ,
\end{equation}
with $\overline{\tilde{J}}_{\alpha,nm} = 0$ for $n = m$. Here, the subscript $\textbf{E}$ denotes a $\textbf{k}$-derivative in the direction of the electric field, and the auxiliary quantities needed for this computation are:

\begin{equation} \label{dAHtoW}
\partial_{\alpha}A_{\textbf{E},nm} = \partial_{\alpha} \left( \overline{\mathbb{A}}_{\textbf{E},nm} + \overline{J}_{\textbf{E},nm} \right)
= \left[ U^{\dagger} \partial_{\alpha}\mathbb{A}_{\textbf{E}} U + i \overline{J}_{\alpha} \overline{\mathbb{A}}_{\textbf{E}} -i \overline{\mathbb{A}}_{\textbf{E}} \overline{J}_{\alpha} + \partial_{\alpha} \overline{J}_{\textbf{E}} \right]_{nm} ,
\end{equation}

\begin{equation} \label{dJHtoW}
\begin{split}
\partial_{\alpha}\overline{J}_{\textbf{E},nm} = & i\partial_{\alpha} \frac{\left[ U^{\dagger}\partial_{\textbf{E}}\mathbb{H} U \right]_{nm}}{\varepsilon_{m\textbf{k}} - \varepsilon_{n\textbf{k}}} \\ = &
\frac{1}{\varepsilon_{n\textbf{k}} - \varepsilon_{m\textbf{k}}}\Big[ - \partial_{\alpha}\left(\varepsilon_{n\textbf{k}} - \varepsilon_{m\textbf{k}}\right) \overline{J}_{\textbf{E}} + \overline{J}_{\alpha} U^{\dagger}\partial_{\textbf{E}}\mathbb{H} U - U^{\dagger}\partial_{\textbf{E}}\mathbb{H} U\overline{J}_{\alpha} - iU^{\dagger}\left(\partial_{\alpha}\partial_{\textbf{E}}\mathbb{H} \right)U \Big]_{nm} .
\end{split}
\end{equation}

We note that in addition to the state-change contribution captured above, there is a contribution to the OHC from the change in band occupation under the electric field. This can be treated within the relaxation time approximation~\cite{atwal2002relaxation} as $\delta f_{n} = e\tau (\partial f_{n}/\partial \varepsilon_{n\textbf{k}}) \textbf{E} \cdot \textbf{v}_{nn}$, where $\tau$ is the relaxation time. This contribution becomes zero in time-reversal symmetric systems and is neglected in the present calculations.

Another way of separating different contributions of the change of orbital current yields

\begin{equation} \label{Current12}
\begin{split}
\langle \delta j_{\alpha}^{L_{\beta}} \rangle^{\textrm{D}} = & \frac{1}{2} \sum_{nm} f_{n}^{\textrm{H}} \left[ \left( \delta v_{\alpha,nm}^{\textrm{D}} L_{\beta,mn}^{\textrm{H}} + L_{\beta,nm}^{\textrm{H}} \delta v_{\alpha,mn}^{\textrm{D}} \right) + \left( v_{\alpha,nm}^{\textrm{H}} \delta L_{\beta,mn}^{\textrm{D}} + \delta L_{\beta,nm}^{\textrm{D}} v_{\alpha,mn}^{\textrm{H}} \right) \right] \\ = & \langle \delta j_{\alpha}^{L_{\beta}} \rangle^{\textrm{D}}_{\textrm{I}} + \langle \delta j_{\alpha}^{L_{\beta}} \rangle^{\textrm{D}}_{\textrm{II}},
\end{split}
\end{equation}
where $\langle \delta j_{\alpha}^{L_{\beta}} \rangle^{\textrm{D}}_{\textrm{I}}$ can be interpreted as the change in velocity, or rate at which the OAM is transported, and $\langle \delta j_{\alpha}^{L_{\beta}} \rangle^{\textrm{D}}_{\textrm{II}}$ as a change in magnitude of the transported OAM with a constant transport velocity. In most of the systems where we have looked at these two contributions were both of large magnitude and opposite sign which hints at two strong effects opposing each other. Future studies of these two contributions could also yield interesting details related to orbital transport.

\section*{E. Computational Parameters}

All computational parameters for the DFT calculations performed with the FLEUR code~\cite{fleurWeb,fleurCode} and the subsequent Wannierization with Wannier90~\cite{pizzi2020wannier90} are summarized in Table~\ref{Table2}. For all materials, the OHC was evaluated on a $250 \times 250 \times 250$ $k$-mesh with a thermal broadening of $\eta = k_{\textrm{B}}T = 0.025\,\mathrm{eV}$, with spin-orbit coupling included self-consistently with the quantization axis along $z$. To suppress poles arising from near-degenerate bands at finite temperature, a small spin Zeeman field of strength $J_{\textrm{Z}} = 10^{-4}\,\mathrm{eV}$ was applied, and contributions from terms containing $\overline{J}$ were discarded when the energy difference between an occupied and an unoccupied band fell below $10^{-4}\,\mathrm{eV}$. Parameters not listed in Table~\ref{Table2} were set to their default values in the respective codes. During Wannierization, the energy region below the Fermi level and a small number of bands above were included in the frozen disentanglement window~\cite{souza2001maximally,marzari1997maximally,marzari2012maximally,pizzi2020wannier90}. The exchange-correlation potential was treated within the generalized gradient approximation using the Perdew-Burke-Ernzerhof (PBE) functional~\cite{blugel2006full,perdew1996generalized}.

\begin{table}[htbp]
\centering
\caption{Input parameters for the FLEUR and Wannier90 calculations. WM denotes the Wannierization mesh and \#PB the number of projected bands. Additional material-specific parameters: lattice constant $c = 4.652$\,\AA\ for hcp Ti; for 2H-MoS$_{2}$, vacuum cell $D_{\textrm{vac}} = 10.31\,a_{0}$ and artificial unit cell $\tilde{D} = 13.96\,a_{0}$ for the LAPW basis construction (upper/lower entries refer to Mo/S atoms respectively). $E_{\textrm{FW}}$ denotes the upper bound of the frozen disentanglement window.}
\arrayrulecolor{black}
\setlength{\arrayrulewidth}{0.3pt}
\renewcommand{\arraystretch}{1.2}
\setlength{\tabcolsep}{3pt}
\resizebox{\textwidth}{!}{%
\begin{tabular}{@{}c|cccccccccccc@{}}
\hhline{-------------}
\rowcolor{gray!15}
\textbf{} & \textbf{a} / \AA & \textbf{K$_{\textrm{max}}$} / $a_{0}^{-1}$ & \textbf{G$_{\textrm{max}}$} / $a_{0}^{-1}$ & \textbf{G$_{\textrm{max}}^{\textrm{XC}}$} / $a_{0}^{-1}$ & \textbf{l$_{\textrm{max}}$} & \textbf{R$_{\textrm{MT}}$} / $a_{0}$ & \textbf{j$_{\textrm{ri}}$} & \textbf{Orbitals} & \#\textbf{PB} & \textbf{DFT mesh} & \textbf{WM} & \textbf{E$_{\textrm{FW}} - $ E$_{\textrm{F}}$} / eV \\
\hhline{-|------------}
\rowcolor{white}
\textbf{fcc Pt} & 3.924 & 5.0 & 15.0 & 12.5 & 12 & 2.3 & 981 & s,p,d & 36 & $16\!\times\!16\!\times\!16$ & $10\!\times\!10\!\times\!10$ & 8.88 \\
\rowcolor{gray!10}
\textbf{bcc W} & 3.190 & 5.0 & 15.0 & 12.5 & 12 & 2.54 & 981 & s,p,d & 42 & $25\!\times\!25\!\times\!25$ & $8\!\times\!8\!\times\!8$ & 8.00 \\
\rowcolor{white}
\textbf{bcc Fe} & 2.834 & 5.2 & 15.0 & 12.5 & 12 & 2.26 & 981 & s,p,d & 30 & $27\!\times\!27\!\times\!27$ & $8\!\times\!8\!\times\!8$ & 6.00 \\
\rowcolor{gray!10}
\textbf{bcc V} & 2.999 & 5.0 & 15.0 & 12.5 & 10 & 2.39 & 981 & s,p,d & 30 & $31\!\times\!31\!\times\!31$ & $10\!\times\!10\!\times\!10$ & 7.00 \\
\rowcolor{white}
\textbf{bcc Cr} & 2.871 & 5.0 & 15.0 & 12.5 & 10 & 2.29 & 981 & s,p,d & 64 & $24\!\times\!24\!\times\!24$ & $8\!\times\!8\!\times\!8$ & 4.00 \\
\rowcolor{gray!10}
\textbf{$\alpha$-Ge} & 5.762 & 5.0 & 15.0 & 12.5 & 12 & 2.29 & 981 & s,p & 32 & $17\!\times\!17\!\times\!17$ & $8\!\times\!8\!\times\!8$ & 2.03 \\
\rowcolor{white}
\textbf{fcc Cu} & 3.637 & 5.0 & 15.0 & 12.5 & 10 & 2.36 & 981 & s,p,d & 36 & $25\!\times\!25\!\times\!25$ & $8\!\times\!8\!\times\!8$ & 7.70 \\
\rowcolor{gray!10}
\textbf{hcp Ti} & 2.937 & 5.0 & 15.0 & 12.5 & 10 & 2.65 & 981 & s,p,d & 72 & $31\!\times\!31\!\times\!17$ & $8\!\times\!8\!\times\!8$ & 3.35 \\
\rowcolor{white}
\textbf{MoS$_2$} & 3.193 & 5.0 & 15.0 & 12.5 & \splittwo{white}{12}{12} & \splittwo{white}{2.43}{1.9} & \splittwo{white}{981}{981} & \splittwo{white}{d}{p} & 22 & $16\!\times\!16\!\times\!1$ & $8\!\times\!8\!\times\!1$ & --- \\
\hhline{-------------}
\end{tabular}%
}
\label{Table2}
\end{table}

\bibliographystyle{naturemag}
\bibliography{reference}